\documentclass[letterpaper]{article}

\usepackage[T1]{fontenc}
\usepackage{url}
\usepackage[numbers,sort&compress]{natbib}

\usepackage{geometry}
\usepackage{setspace}
\usepackage[normalem]{ulem}

\usepackage{graphicx}
\usepackage{caption}
\usepackage{subcaption}
\usepackage{float}
\newfloat{scheme}{htbp}{los}
\floatname{scheme}{Scheme}
\floatname{chart}{Chart}
\newfloat{graph}{htbp}{loh}

\usepackage{chemformula} % Formulas using \ch{}
\usepackage[version = 4]{mhchem} % Formulas using \ce{}
\usepackage[colorlinks=true,
linkcolor=blue,
citecolor=blue,
urlcolor=blue]{hyperref}

\usepackage{authblk}
\author[1]{Afreen Anamul Haque}
\author[1]{Aniket Singha}
\affil[1]{Department of E \& ECE, IIT Kharagpur, India}

\title{Transition-Metal Tailored Ga\textsubscript{2}O\textsubscript{2} Monolayer: From Room-Temperature Gas Sensing to Chemical Scavenging}%Dopant-Tuned Ga\textsubscript{2}O\textsubscript{2} Monolayers for Selective Gas Detection and Chemical Scavenging
\date{*Email:afreenhaque28@gmail.com}

\begin{document}

\maketitle

\begin{abstract}

  Pristine Ga\textsubscript{2}O\textsubscript{2} monolayers suffer from poor sensitivity and weak molecular capture, limiting their application in toxic gas detection and environmental detoxification. Here, we employ first-principles density functional theory (DFT) calculations  to investigate the gas sensing and scavenging properties of Ga\textsubscript{2}O\textsubscript{2} monolayers substitutionally tailored via seven transition-metals (TM): Pd, Zn, Zr, Mo, Ag, Ti, and Pt. All TM-substituted monolayers exhibit negative formation and binding energies, negligible lattice distortion, and structural stability in molecular dynamics simulations. Performance evaluation against eight toxic industrial and three environmental gases reveals functionalities ranging from selective, reusable room-temperature sensing to permanent molecular capture. Ag substitution exhibits exceptional selectivity for NO with moderate adsorption strength ($\sim-0.83eV$), an up to eight-order-of-magnitude conductivity enhancement, besides facilitating reusable O\textsubscript{2} and NO\textsubscript{2} detection. Additionally, Pd, Zn, Zr, and Mo substitutions tune selectivity toward NO, NO\textsubscript{2}, CO\textsubscript{2}, CO, and O\textsubscript{2}. Coming to applications towards toxic gas capture, Zr- and Mo-substituted systems selectively scavenge oxidizing gases, whereas Ti and Pt act as universal scavengers. Further analysis reveals that, Pd- and Ag-substituted monolayers remain selective for NO, while Zn substitution favors NO\textsubscript{2} detection even in ambient atmospheric conditions. Thus, these tailored Ga\textsubscript{2}O\textsubscript{2} monolayers offer a practical platform for atmospheric monitoring and detoxification.
  
\end{abstract}

\section*{Keywords}
2D materials based gas sensors, Density Functional Theory (DFT), Novel Ga\textsubscript{2}O\textsubscript{2}-based gas sensors, substiution, molecule scavenging

%\section*{Abbreviations}

%Some journals require a list of abbreviations: these normally should be given immediately after the keyswords (if required).

%%%%%%%%%%%%%%%%%%%%%%%%%%%%%%%%%%%%%%%%%%%%%%%%%%%%%%%%%%%%%%%%%%%%%
%% Start the main part of the manuscript here.
%%%%%%%%%%%%%%%%%%%%%%%%%%%%%%%%%%%%%%%%%%%%%%%%%%%%%%%%%%%%%%%%%%%%%
\section{Introduction}

The detection and capture of hazardous gas molecules at trace concentrations is a challenge of growing urgency across environmental monitoring, industrial safety, and biomedical diagnostics. Toxic species such as NO, NO\textsubscript{2}, SO\textsubscript{2}, NH\textsubscript{3}, H\textsubscript{2}S, CO, CS\textsubscript{2} etc. are implicated in urban air pollution, respiratory disease, and industrial process hazards and pose critical risks in confined environments~\cite{zhao2020gas, zhang2019recent}. Additionally, HF, a highly corrosive compound widely encountered in semiconductor fabrication and chemical processing industries, poses severe toxicity hazards even at trace concentrations. Hence, the detection of these molecules has become quite imperative.\\

\begin{figure}[htbp]
	\centering
	\includegraphics[width=0.8\columnwidth]{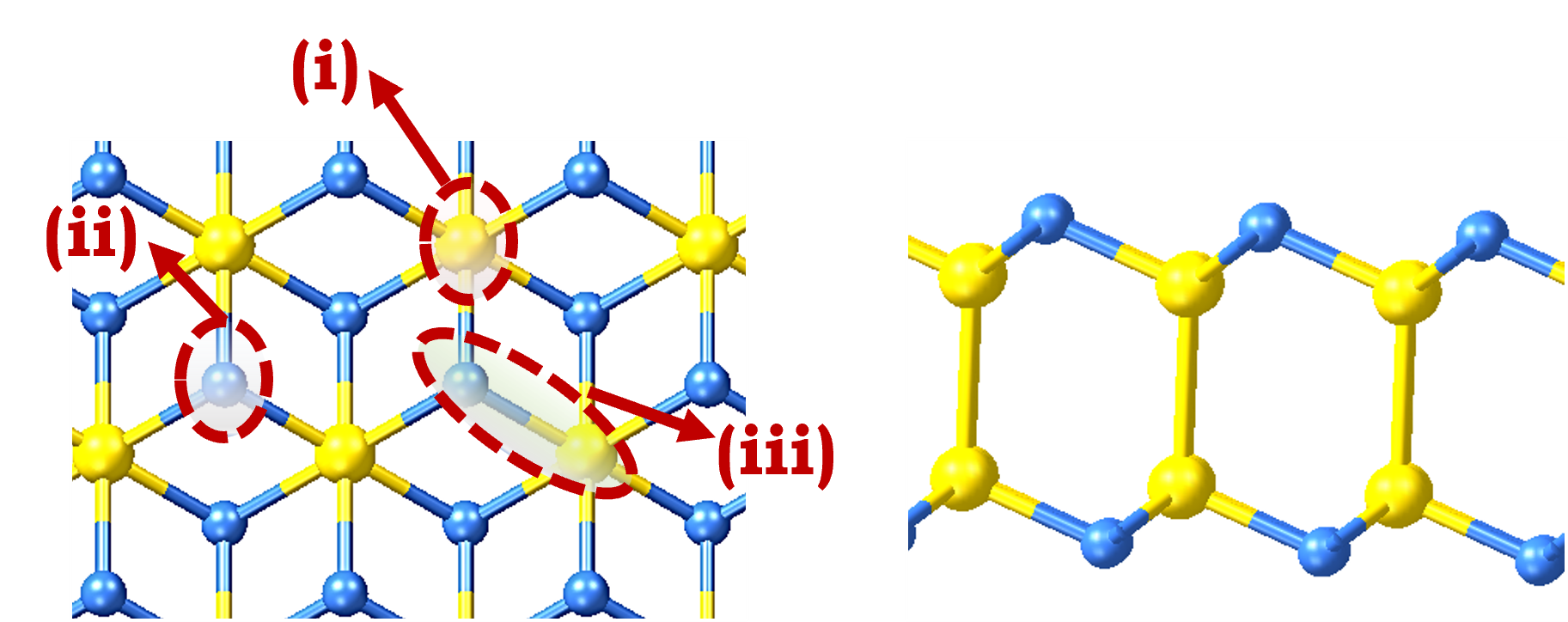}
	\caption{Left Panel: Top view of the 2D Ga\textsubscript{2}O\textsubscript{2} monolayer along with the schematic representation of the three sites considered for transition-metal substitution: (i) Ga-site substitution, where the TM replaces a Ga atom; (ii) O-site substitution, where the TM replaces an O atom; and (iii) Stone–Wales (SW) defect-site substitution, where the TM occupies the site created by the Stone–Wales defect. The location of the subsitution sites have been encircled in red color. Right Panel: Side view of the pristine Ga\textsubscript{2}O\textsubscript{2} monolayer.}
	\label{fig:config_positions}
\end{figure}

\noindent Recently, two-dimensional (2D) materials have emerged as compelling candidates for next-generation gas sensing platforms owing to their high surface-to-volume ratio, tunable electronic properties, and compatibility with miniaturized device architectures~\cite{buckley2020frontiers, wang2022gas, vincent2021opportunities, chaves2020bandgap, liu2021, Huyghebaert2018}. A wide range of 2D materials including graphene~\cite{buckley2020frontiers, schedin2007detection}, transition metal dichalcogenides~\cite{szary2026first}, hexagonal boron nitride~\cite{kalwar2022highly, ahmad2025first, wu2013mechanics}, and metal oxide monolayers~\cite{zhao2021two, al2026, zhang2024ultrafast, haque2025gas} etc., are being explored for sensing applications. The particular class of 2D-metal oxides have attracted growing attention due to their intrinsic semiconducting character and rich surface chemistry, which enable diverse molecule–surface interactions~\cite{zhao2021two}. Unlike conventional bulk metal oxide gas sensors, which often require elevated operating temperatures because of their wide bandgaps, 2D metal oxides offer the possibility of lower temperature operation due to their reduced bandgaps, compared to their bulk counterparts~\cite{hahn2010metal, nano8100851}. All these advantages further make them promising candidates for room-temperature gas sensing applications. Among the emerging members of this family, a recently proposed two-dimensional phase of Gallium Oxide with stoichiometric formula Ga\textsubscript{2}O\textsubscript{2} has attracted growing interest owing to its structural stability, oxidation resistance, moderate electronic bandgap of $2.76eV$ (HSE06) and high hole mobility~\cite{shao2021two, Demirci2017, Li2026, Yang2026, alfaqiri2023, Liu2026}. The Ga\textsubscript{2}O\textsubscript{2} monolayer (ML) adopts a buckled, graphene-like structure stabilized by strong directional Ga-O bonding~(see Fig.~S1 in Supporting Information (SI)). This moderate bandgap may facilitate adsorption-induced conductivity modulation at room temperature, making the Ga\textsubscript{2}O\textsubscript{2} ML a promising candidate to study for gas sensing applications. In a prior investigation, we established the intrinsic gas sensing characteristics of the pristine Ga\textsubscript{2}O\textsubscript{2} ML, demonstrating selective detection of NH\textsubscript{3} at room temperature as its primary sensing capability, with NO detection accessible only under low-temperature conditions~\cite{haque2025predictive}. However, the adsorption energy for most other hazardous gas species remain modest in the pristine ML. This hampers the possibility of  detecting molecules other than NH\textsubscript{3} and also precludes irreversible molecular capture limiting the material’s broader sensing and scavenging potential.\\

\noindent Substitutional doping with transition metal (TM) atoms offers a direct and chemically versatile route to overcome these limitations~\cite{Zhang_Robinson_2019, Han2019, panigrahi2019, Maity2024}. By introducing dopant-specific d-states in the bandgap, TM substituents fundamentally reshape the electronic landscape of the host ML, enabling interactions with target analytes beyond those facilitated by the pristine surface~\cite{petterson2014}. Transition metals spanning the periodic table differ systematically in their d-band occupation, electronegativity, and atomic radius, each introducing a distinct local electronic and structural perturbation to the Ga\textsubscript{2}O\textsubscript{2} lattice. Early transition metals such as Ti (Z=22)~\cite{ahmad2025first, kalwar2022highly} and Zr (Z=40)~\cite{guo2022zr, abdesselem2026hazardous}, together with middle- and late-series dopants including Zn (Z=30)~\cite{su2023effect, mollaamin2023transition}, Mo (Z=42)~\cite{cheghib2026first}, Pd (Z=46)~\cite{akter2025hazardous, wang2019characterization, viveka2025}, Ag (Z=47)~\cite{guo2022dft, zhang2026dft}, and Pt (Z=78)~\cite{kong2024metal, wang2019characterization, chen2018noble}, possess chemically accessible d states that facilitate strong adsorbate–surface hybridization and charge redistribution, thereby profoundly influencing molecular adsorption characteristics. This systematic progression across the transition metal series provides a rational basis for exploring a wide functional landscape, from selective room-temperature sensing to irreversible molecular scavenging. In this work, a comprehensive adsorption and sensing investigation of the TM substituted Ga\textsubscript{2}O\textsubscript{2} ML is carried out across eleven gas species, comprising eight hazardous analytes: NH\textsubscript{3}, NO, SO\textsubscript{2}, NO\textsubscript{2}, H\textsubscript{2}S, CO, CS\textsubscript{2}. To gauge the possibility of deploying in ambient environment, the interaction of the substituted ML with three ambient atmospheric gases such as O\textsubscript{2}, CO\textsubscript{2}, and H\textsubscript{2}O are also explored. Owing to the chemically inert nature of N\textsubscript{2}, its interaction with the surface is not examined in detail. \\
%Early transition metals such as Ti (Z=22)~\cite{ahmad2025first, kalwar2022highly} and Zr (Z=40)~\cite{guo2022zr, abdesselem2026hazardous}, owing to their strong Lewis acidic character and chemically accessible d states, are expected to promote pronounced adsorbate–surface hybridization and charge redistribution interactions. Middle-series dopants such as Zn (Z=30)~\cite{su2023effect, mollaamin2023transition} and Mo (Z=42)~\cite{cheghib2026first} introduce distinct electronic hybridization and redox characteristics, thereby modulating the adsorption behavior and charge redistribution at the surface. Noble and late transition-metal dopants including Pd (Z=46)~\cite{akter2025hazardous, wang2019characterization, viveka2025}, Ag (Z=47)~\cite{guo2022dft, zhang2026dft} and Pt (Z=78)~\cite{kong2024metal, wang2019characterization, chen2018noble} possess electronically accessible d states that promote substantial adsorbate-surface hybridization, thereby strongly influencing molecular adsorption characteristics.

\noindent This paper is organized as follows. In Sec.~\ref{sec:method}, we highlight the details of our computational methodology and the specifics of the parameters employed for calculations.  Next, in Sec.~\ref{result}, we discuss the gas sensing performance of the substituted Ga\textsubscript{2}O\textsubscript{2} ML, where we first discuss the adsorption properties (Sec.~\ref{subsec:adsorption}) and then proceed towards discussing its performance as a reusable gas sensor and as a scavenger. We conclude the paper briefly in Sec.~\ref{conclusion}. 

\section{Computational Methods} 
\label{sec:method}

\indent First-principles density functional theory (DFT) calculations were performed using the Vienna \textit{ab initio} Simulation Package (VASP). The exchange-correlation interactions were treated within the generalized gradient approximation (GGA) using the Perdew-Burke-Ernzerhof (PBE) functional in the projector augmented wave (PAW) framework~\cite{perdew1996generalized, kresse1996efficiency, blochl1994projector}. For Ga, the semicore d-states were treated as valence states to properly assess bonding with the neighbouring atoms and adsorption of gas molecules. The substituted metal atoms were described using the corresponding PAW pseudo-potentials, with semicore p states included in the valence configuration where available. A $4 \times 4 \times 1$ supercell was constructed to study the adsorption characteristics of the gas molecules. Structural optimization of the supercell was carried out using a $3 \times 3 \times 1$ Monkhorst-Pack k-point mesh and a plane-wave cutoff energy of $500eV$. Gaussian smearing with a width of 0.05eV was employed to facilitate electronic convergence during structural relaxation and self-consistent field (SCF) calculations. For the density of states (DOS) calculations, the tetrahedron method was used for Brillouin-zone integration to obtain accurate electronic occupations and DOS features. A vacuum layer of approximately 25\AA~was introduced along the out-of-plane direction to avoid spurious interlayer interactions arising from periodic boundary conditions. The van der Waals interactions were incorporated using Grimme’s DFT-D3 method with zero damping~\cite{goerigk2017comprehensive, witte2017assessing}. All calculations were considered converged when the total energy difference between successive self-consistent field iterations was below $10^{-8}$eV and the maximum residual force on each atom was less than $10^{-4}$eV/\AA. \\

\begin{table}[ht]
	\centering
	\caption{Calculated values of formation energy ($E_{\text{f}}$) and binding energy ($E_{\text{bind}}$) of the most energetically favorable configuration for all the seven substituted monolayers under consideration.}
	%\resizebox{\columnwidth}{!}{
		\begin{tabular}{lccc}
			\hline
			Element & Formation Energy & Binding Energy & Environmental \\
			& $E_{\text{f}}$, (eV) & $E_{\text{bind}}$, (eV) & Conditions \\
			\hline
			Ti & -3.92  & -4.83 & O-rich  \\
			Zn & -5.05 & -9.31 & O-rich   \\
			Zr & -6.86 & -17.63 & O-rich  \\
			Mo & -2.27 & -16.64 & O-rich  \\
			Pd & -2.31 & -9.16 & O-rich  \\
			Ag & -3.29 & -9.11 & O-rich  \\
			Pt & -1.16 & -5.35 & O-rich  \\
			\hline
		\end{tabular}
%	}
	\label{tab:formation_binding}
\end{table}

\noindent The converged in-plane lattice constant of the pristine supercell was found to be 12.47~\AA. To determine the preferred substitution site for a TM atom, we modeled three cases in the Ga\textsubscript{2}O\textsubscript{2} monolayer: (i) TM substituting a Gallium atom (Ga vacancy), (ii) TM substituting an Oxygen atom (O vacancy), and (iii) TM incorporated at a Stone–Wales (SW) defect, as demonstrated in Fig.~\ref{fig:config_positions}. Each dopant atom (Ti, Zn, Zr, Mo, Pd, Ag, and Pt) was substituted at these defect sites followed by full structural relaxation~(see Fig.~S2 in SI). Out of the three cases, the configuration with the lowest formation energy is most likely to form spontaneously. Hence, this configuration was selected for further investigation of gas sensing capabilities. The values of the formation energy for the most probable configuration in each case is given in Table~\ref{tab:formation_binding}. In addition to the formation energy, the thermodynamic stability of the most stable configuration of each of the seven substituted systems was evaluated using the calculated binding energy and molecular dynamics (MD) simulations. For gas adsorption studies, the gas molecules were initially positioned at a vertical distance of 2.5\AA~above the ML surface at the site of TM substitution. For each gas molecule, three initial adsorption geometries, namely; one parallel configuration and two perpendicular configurations to the plane of ML were examined. In the perpendicular arrangements, each terminal atom of polar molecules was separately oriented toward the substituted TM site to account for orientation-dependent adsorption effects due to dipole formation. The relaxed configurations for all the adsorption cases under investigation has been included in SI.\\
%three different initial adsorption geometries were considered for each gas molecule. These orientations are as follows: a molecule (i) horizontally or parallel, (ii) vertically or perpendicular, to the plane of the ML. For the vertical orientation, in case of polar molecule, computations were done while placing both the ends of the molecule close to the substituted TM atom. 

\section{Results and Discussion}
\label{result}

As stated in the last section, the relative formation probabilities of all the substituted configurations under investigation were evaluated through the calculation of the corresponding formation energies ($E_{f}$). Formation energy is defined as the energy required to form a defect or substitutional configuration in a material relative to the pristine structure. A negative formation energy indicates that the defect or substitution is likely to form spontaneously. For a defect configuration, $E_f$ is calculated as: 

\begin{equation}
	\label{eq:formation_energy}
	E_{f} = E_{tot}^{sub} - \mu_{sub} + m\mu_{Ga} + n\mu_{O} - E_{tot}^{PL},
\end{equation}

\noindent where $E_{tot}^{sub}$ is the total energy of the TM atom-substituted Ga\textsubscript{2}O\textsubscript{2} ML, $E_{tot}^{PL}$ is the total energy of the pristine ML and $\mu_{sub}$ corresponds to the average chemical potential of the substitutional atom in the most stable bulk phase. Here,  \textit{m} and \textit{n} represent the number of Ga and O atoms removed, respectively. The chemical potentials of $\mu_{Ga}$ and $\mu_{O}$ under Ga-rich and O-rich conditions, respectively, were obtained from the average energies of the corresponding most stable reference phases of these elements under room temperature and normal atmospheric pressure. On the other hand, the values for $\mu_{Ga}$ and $\mu_{O}$ for O-rich and Ga-rich conditions, respectively, were calculated using the following expression~\cite{van2004first, zhang1991chemical, rogal2014perspectives} :
\begin{subequations}
	\label{eq:mu_Ga}
	\begin{align}
		4\mu_{\mathrm{Ga}} &= 2\mu_{\mathrm{Ga_{2}O_{3}}} - 3\mu_{\mathrm{O_{2}}}, \label{eq:mu_Ga_2} \\
		3\mu_{\mathrm{O}} &= \mu_{\mathrm{Ga_{2}O_{3}}} - 2\mu_{\mathrm{Ga}}, \label{eq:mu_Ga_3}
	\end{align}
\end{subequations}
%Under Ga-rich conditions, $\mu_O$ is determined by the $\mu_{Ga}$ in bulk and the $\mu$ of the most stable bulk phase of the oxide, i.e. the $\beta$-Ga\textsubscript{2}O\textsubscript{3} using the following expression:
%calculated as the total DFT energy of the bulk unit cell divided by the number of atoms in the unit cell.  The chemical potentials of Ga and O are constrained by phase stability with respect to the reference oxide phase i.e the $\beta$-Ga\textsubscript{2}O\textsubscript{3}, as defined in Eq.~\ref{eq:mu_Ga}. These relations ensure that the chosen values of $\mu_{\mathrm{Ga}}$ and $\mu_{\mathrm{O}}$ do not lead to the formation of  other competing phases.

%Two limiting conditions are considered, namely Ga-rich and O-rich environments, yielding four possible cases for $\mu_{\mathrm{Ga}}$ and $\mu_{\mathrm{O}}$. Under Ga-rich conditions, $\mu_{\mathrm{Ga}}$ attains its maximum value corresponding to bulk Ga, while $\mu_{\mathrm{O}}$ is reduced accordingly. In contrast, under O-rich conditions, $\mSIu_{\mathrm{O}}$ reaches its upper bound given by $\tfrac{1}{2}\mu_{\mathrm{O_2}}$, and $\mu_{\mathrm{Ga}}$ decreases correspondingly. These limits define the range of chemical potentials used in the formation energy calculations.
\noindent In the above equation, $\mu_{Ga_{2}O_{3}}$ represents the average chemical potential of $\beta$-Ga\textsubscript{2}O\textsubscript{3}, which is the most stable oxide of Gallium~\cite{pearton2018, tsao2018, higashiwaki2018}. Among the considered configurations, TM substitution at the Ga site under O-rich conditions was found to exhibit the lowest formation energy and thus, is considered as energetically favourable. In all the cases, the energetically favourable substitution configuration demonstrated negative formation energy (Table \ref{tab:formation_binding}) hinting towards the possibility of spontaneous formation under O-rich conditions. In addition, the binding energy ($E_{\text{bind}}$) was calculated using the expression:
%The binding energy provides insight into the energetic stability of the dopant within the ML, where a more negative value indicates stronger binding and a more stable substituted configuration. 

\begin{equation}
	E_{bind} = E_{\text{tot}}^{\text{sub}} - E_{\text{tot}}^{\text{vac}} - \mu_{\text{sub}},
\end{equation}

\noindent where $E_{\text{tot}}^{\text{vac}}$ is the total energy of the ML containing the corresponding vacancy. All the TM substitutions under consideration display negative binding energies, indicating attractive interaction between the TM atom and the host lattice, thereby suggesting structural stability. In addition, the average change in the in-plane lattice constant in each of these cases was noted to be less than 1\%, highlighting negligible structural distortion on substitution with the TM atoms. To evaluate the finite-temperature thermodynamic stability of the substituted Ga\textsubscript{2}O\textsubscript{2} monolayer, Ab-initio Molecular Dynamics (AIMD) simulations were carried out for all seven dopant configurations at $1000~K$. In all cases, the energy fluctuations remain bounded without noticeable systematic drift, and no structural degradation or dopant migration is observed in any of the simulation, indicating good dynamical stability of the substituted systems. The corresponding AIMD plots depicting temporal energy fluctuation at $T=1000K$ are included in the Fig.~S3 in SI.

\subsection {Adsorption Characteristics}
\label{subsec:adsorption}

To gauge the possibility of reasonable interaction of the gas molecules with TM-substituted ML, the adsorption energy $(E_{ads})$ was calculated by using the expression given below:
\begin{equation}
	E_{ads}=  E_{\text{tot}}^{\text{sub + molecule}} - E_{\text{tot}}^{\text{molecule}} -  E_{\text{tot}}^{\text{sub}},
\end{equation}
\noindent where $E_{\text{tot}}^{\text{sub + molecule}}$ is the total energy of the substiuted system alongwith the molecule and $E_{\text{tot}}^{\text{molecule}}$ is the total energy of the isolated molecule. To classify stable adsorption at and around room temperature, a lower bound of $-0.4eV$ was imposed on the adsorption energy~\cite{han2023theoretical, zhao2021two, wang2014first, Agboobla2021}.
%This threshold is physically motivated by the average thermal kinetic energy of a gas molecule at room temperature at T = 300 K, this amounts to 3/2$k_B$T=0.04. A negative adsorption energy roughly one order of magnitude larger than this value (~0.4 eV) is generally sufficient to retain the molecule on the surface long enough to register a time-averaged, electrically measurable signal; a criterion well-established in the two-dimensional materials sensing literature.
Adsorption energies weaker than (less negative) $-0.4eV$ implies that thermal fluctuations under ambient conditions may result in rapid desorption with recovery times too short for reliable electrical detection. In this work, adsorption energies in the range of $-0.4eV$ to $-1eV$ are considered optimal for reusable gas sensing, as they provide a balance between stable adsorption and spontaneous surface regeneration under ambient conditions. Adsorption energies between $-1eV$ and $-1.5eV$ remain potentially useful for gas sensing; however, surface regeneration may necessitate thermal annealing or UV-assisted desorption. When adsorption energies surpass $-1.5eV$, the interaction is effectively irreversible under practical operating conditions, rendering the material more suitable as a one time sensor or molecular gas scavenger rather than a reusable detection platform. Accordingly, the adsorption energy window of $-0.4eV$ to $-1.5eV$ is adopted here as the favorable regime for gas capture, with the sub-ranges distinguished according to their expected recovery characteristics as explained earlier. In addition to adsorption energy, the adsorption height (H) and net charge transfer (Q) between the adsorbate and adsorbent also provide valuable insights into the nature of adsorption. Adsorption height refers to the perpendicular distance between the parent layer and the molecule in the most stable configuration. To achieve stable and robust adsorption, a reasonable level of charge transfer between the substrate and the molecules is essential. Bader Partition Algorithm~\cite{bader1985atoms} was employed to compute the charge distribution and assess the charge transfer between the substituted Ga\textsubscript{2}O\textsubscript{2} ML and the gas molecule and was calculated using the expression Eq.~S1 in SI. In order to assess the reusability of the gas sensors, we have also computed the recovery time ($\tau$) using Arrhenius equation, Eq.~S2 given in SI. We note that the recovery time should be of the order of milliseconds to few seconds for both suitable electrical detection and spontaneous surface regeneration in a reusable gas sensor.\\

%\begin{figure}[htbp]
%	\centering
%	\includegraphics[width=0.9\columnwidth]{Pictures/Pd_substitution.png}
%	\vspace{0.05em}
%	\includegraphics[width=0.9\columnwidth]{Pictures/Ag_substitution.png}
%	\caption{(a) Pd substitution. (b) Ag substitution.}
%	\label{fig:bubble}
%\end{figure}

%\begin{figure*}[htbp]
%	\centering
%	\begin{subfigure}{0.9\textwidth}
%		\centering
%		\includegraphics[width=\linewidth]{Pictures/Pd_substitution.png}
%		\caption{Pd-Ga\textsubscript{2}O\textsubscript{2} ML}
%		\label{fig:Pd_bubble}
%	\end{subfigure}
%		\vspace{0.5em}
%	\begin{subfigure}{0.9\textwidth}
%		\centering
%		\includegraphics[width=\linewidth]{Pictures/Ag_substitution.png}
%		\caption{Ag-Ga\textsubscript{2}O\textsubscript{2} ML}
%		\label{fig:Ag_bubble}
%	\end{subfigure}
%	\caption{Bubble graph showing charge transfer (e), Recovery time (s), and adsorption height (Å) for TM-substituted ML. The size of the bubbles is indicative of the magnitude of the adsorption energy (eV).}
%	\label{fig:bubbles}
%\end{figure*}

\noindent While these descriptors provide a comprehensive picture of the strength, stability, and reversibility of adsorption, they do not directly reveal whether adsorption can induce a measurable electrical response. Since the operating principle of a resistive gas sensor relies on adsorption-induced modifications to the electronic structure of the sensing layer, a detailed analysis of the electronic density of states (DOS) is required to establish the sensing capability of the material toward a given analyte. Therefore, in addition to adsorption attributes, we systematically examine the modification of the electronic DOS upon exposure to each target gas molecule. Adsorption of target molecules might induce electronic perturbations such as gap state formation, Fermi level shifts, and changes in the effective bandgap of the adsorbate-ML system which often reflects as a change in the conductance of the system. These features reveal the donor or acceptor character of each adsorbate and its effect on the mobile carrier population. Together, they provide a rigorous electronic basis for interpreting the expected changes in electrical conductivity upon adsorption. Throughout this paper, electron excitation energy refers to the minimum energy required for an electron to transition from a completely or partially filled state to the conduction band edge (CBE). Analogously, hole excitation energy refers to the minimum energy required to excite a hole in the valence band (VB) and is given by the energy difference between the valence band edge (VBE) and the closest completely or partially unoccupied state. \\

\noindent In what follows, we present a systematic analysis of the interaction between gas molecules and TM-substituted Ga\textsubscript{2}O\textsubscript{2} ML for each of the seven TM species. We evaluate the adsorption characteristics, electronic structure modifications, and conductometric response for all of the analytes under consideration. Based on the adsorption energy, recovery time, charge transfer and modulation in DOS, each substitution case is assessed for its suitability as a reusable resistive gas sensor, or a  molecular scavenger or none,  establishing a comprehensive dopant-dependent functional classification of the substituted Ga\textsubscript{2}O\textsubscript{2} ML.

\subsection{Pd Substitution}
\label{subsec:Pd_sub}

\begin{figure*}[htbp]
	\centering
	
	\begin{minipage}[c]{0.50\textwidth}
		\centering
		\scriptsize
		\setlength{\tabcolsep}{6pt}
		\renewcommand{\arraystretch}{1.3}
		
		\begin{tabular}{lcccc}
			\hline
			Molecule & $E_{ads}$ (eV) & H (\AA) & q (e) & $\tau$ (s) \\
			\hline
			NH$_3$   & -0.83 & 1.75 & 0.19  & $1.90 \times 10^{2}$ \\
			NO       & -1.20 & 1.22 & 0.39  & $1.44 \times 10^{8}$ \\
			SO$_2$   & -0.72 & 2.52 & 0.03  & $1.25$ \\
			O$_2$    & -0.28 & 1.89 & 0.067 & $5.06 \times 10^{-8}$ \\
			NO$_2$   & -0.65 & 1.51 & 0.21  & $8.31 \times 10^{-2}$ \\
			H$_2$S   & -0.73 & 0.99 & 0.004 & $1.83$ \\
			H$_2$O   & -0.63 & 0.62 & 0.03  & $3.83 \times 10^{-2}$ \\
			CO       & -0.16 & 2.59 & 0.005 & $4.87 \times 10^{-10}$ \\
			CO$_2$   & -0.57 & 2.27 & 0.02  & $3.76 \times 10^{-3}$ \\
			CS$_2$   & -0.75 & 2.58 & 0.008 & $3.98$ \\
			HF       & -0.31 & 2.29 & 0.03  & $1.61 \times 10^{-7}$ \\
			\hline
		\end{tabular}
		
		\vspace{2.0em}
		{\normalsize\textbf{(a)}}
		\label{tab:Pd_sub}
	\end{minipage}
	\hfill
	\begin{minipage}[c]{0.48\textwidth}
		\centering
		\includegraphics[width=0.95\linewidth]{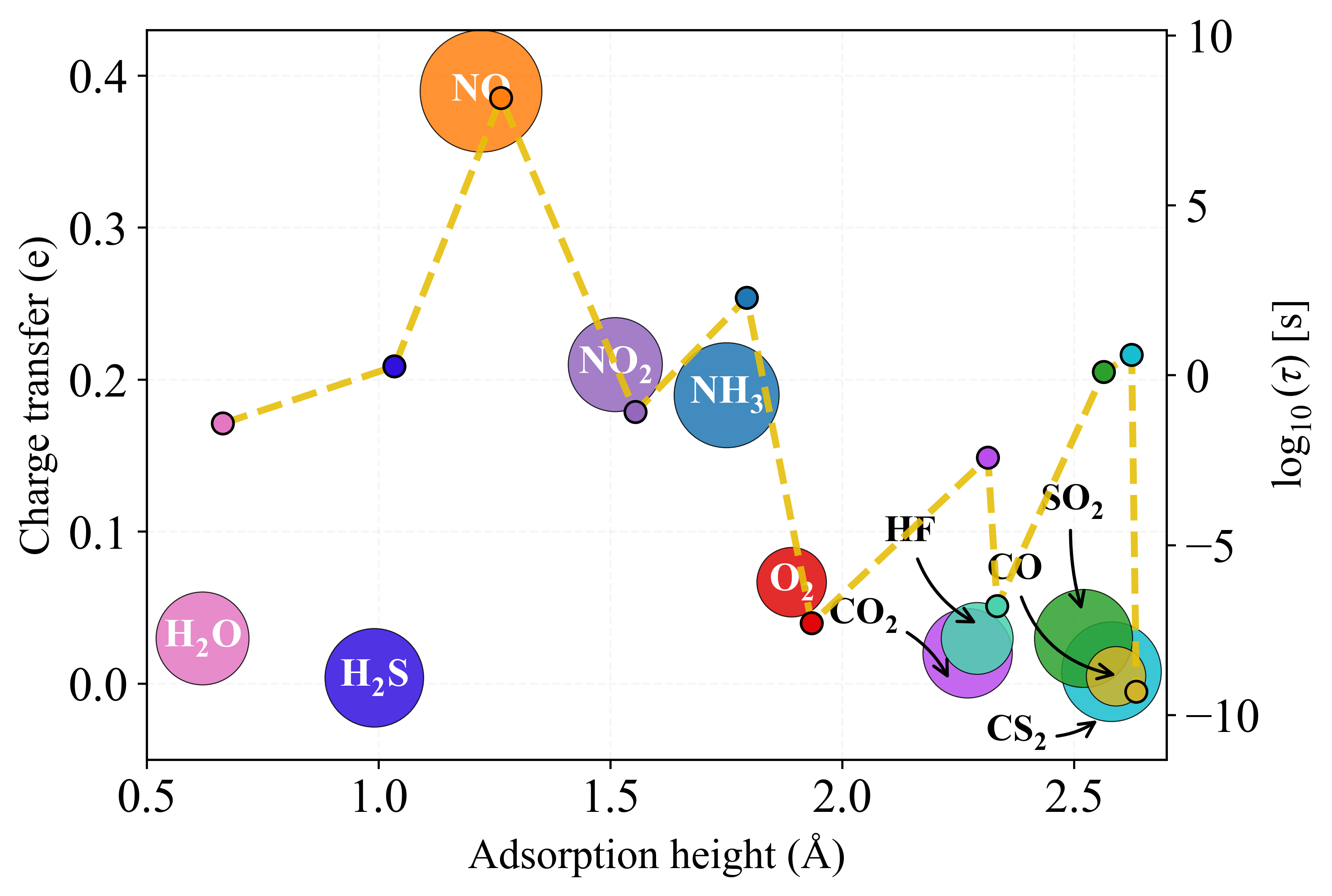}
		
		\vspace{0.3em}
		\textbf{(b)}
	\end{minipage}
	
	\caption{(a) Table I: Adsorption parameters of the investigated gas molecules on the Pd-Ga\textsubscript{2}O\textsubscript{2} ML, including adsorption energy ($E_{ads}$), adsorption height (H), charge transfer (q), and recovery time ($\tau$). (b) Graphical summary of the adsorption characteristics of gas molecules on the Pd-Ga\textsubscript{2}O\textsubscript{2} monolayer. The centers of the larger bubbles are referenced to the X and primary Y-axes, representing the adsorption height and charge transfer, respectively. The area of each bubble is proportional to the magnitude of the adsorption energy. The smaller markers linked by the dashed line indicate the corresponding recovery times, referenced to the secondary Y-axis. Recovery times are reported on a logarithmic scale.}
	\label{fig:Pd_summary}
\end{figure*}

Pd substitution produces a broad spectrum of adsorption responses across the investigated analytes. The corresponding values of $E_{ads}$, Q, H, and $\tau$ for all the considered gas molecules on Pd-substituted ML (Pd-Ga\textsubscript{2}O\textsubscript{2}) are summarized in Fig.~\ref{fig:Pd_summary}. Molecules such as O\textsubscript{2} ($E_{ads}$=$-0.28eV$), CO ($-0.16eV$) and HF ($-0.31eV$) display adsorption energies above (greater than) the $-0.4eV$ threshold, characteristic of weak adsorption behaviour and insufficient for stable signal detection. Accordingly, their DOS hasn't been investigated in detail. A different adsorption regime was observed for molecules like NH\textsubscript{3} ($-0.85eV$), SO\textsubscript{2} ($-0.72eV$), NO\textsubscript{2} ($-0.65eV$), H\textsubscript{2}S ($-0.73eV$), CO\textsubscript{2} ($-0.57eV$), H\textsubscript{2}O ($-0.63eV$), and CS\textsubscript{2} ($-0.75eV$). They exhibited adsorption energies within the favorable range for reusable sensing. \\
%\vspace{-0.2cm}
%\begin{table}[ht]
%	\centering
%	\caption{Values of adsorption energy (E\textsubscript{ads}), charge transfer (q), recovery time ($\tau$), and adsorption height (H) for different molecules investigated on the Pd–Ga\textsubscript{2}O\textsubscript{2} ML.}
%	\begin{tabular}{lcccc}
%		\hline
%		\textbf{Molecule} & \textbf{$E_{ads}$} (eV) & \textbf{H}(~\AA{}) & \textbf{q}(e) & \textbf{$\tau$}(s) \\
%		\hline
%		NH\textsubscript{3}  & -0.83  & 1.75  & 0.19  & $1.90 \times 10^{2}$ \\
%		NO                   & -1.20 & 1.22 & 0.39 & $1.44 \times 10^{8}$ \\
%		SO\textsubscript{2}  & -0.72 & 2.52 & 0.03 & $1.25$ \\
%		O\textsubscript{2}  & -0.28 & 1.89 & 0.067 & $5.06 \times 10^{-8}$ \\
%		NO\textsubscript{2}  & -0.65 & 1.51 & 0.21 & $8.31 \times 10^{-2}$ \\
%		H\textsubscript{2}S& -0.73 & 0.99 & 0.004 & $1.83$ \\
%		H\textsubscript{2}O& -0.63 & 0.62 & 0.03 & $3.83 \times 10^{-2}$ \\
%		CO                 & -0.16 & 2.59 & 0.005 & $4.87 \times 10^{-10}$ \\
%		CO\textsubscript{2}& -0.57 & 2.27 & 0.02 & $3.76 \times 10^{-3}$ \\
%		CS\textsubscript{2}& -0.75 & 2.58 & 0.008 & $3.98$ \\
%		HF                 & -0.31 & 2.29 & 0.03 & $1.61 \times 10^{-7}$ \\
%		\hline
%	\end{tabular}
%	\label{tab:pd_adsorption_values_1}
%\end{table}

\begin{figure*}[!htbp]
	\centering
	\includegraphics[width=\textwidth]{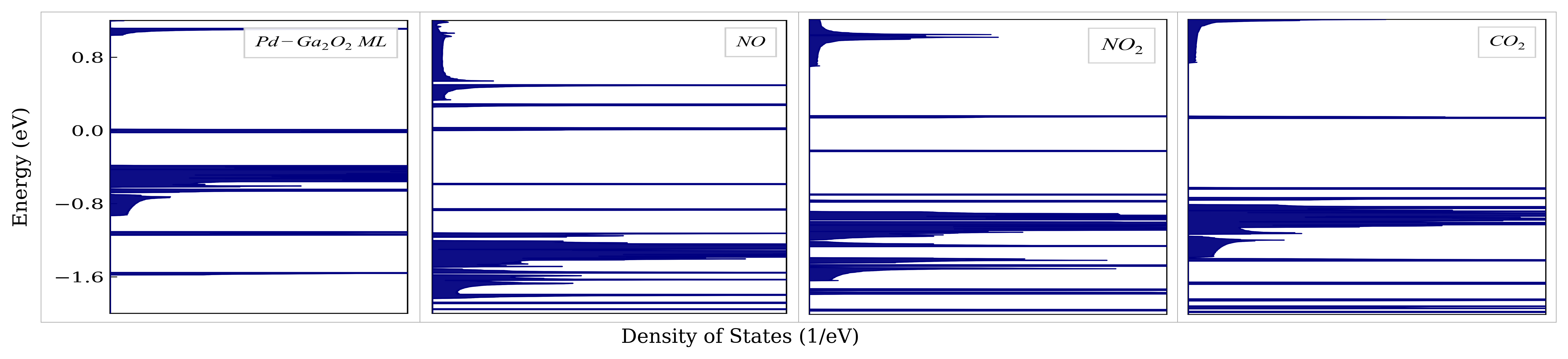}
	\caption{Total DOS of the Pd-Ga\textsubscript{2}O\textsubscript{2} ML and upon adsorption of  the favorable target molecules. The Fermi level is pinned at $0eV$ in all the panels.}
	\label{fig:DOS_Pd}
\end{figure*}

\noindent DOS analysis of the Pd–Ga\textsubscript{2}O\textsubscript{2} ML reveals a localised partially filled state at the Fermi level, with contributions dominated by the Pd atom and its immediate coordination environment (see Fig.~\ref{fig:DOS_Pd}). This state, located approximately $0.35eV$ above the VBE, imparts an elevated baseline p-type conductivity or the doping effect, against which adsorption-induced conductivity modulations must be assessed. It is noteworthy that oxides of Gallium are commonly reported to exhibit intrinsic n-type behavior due to inherent donor impurities introduced during synthesis or fabrication processes~\cite{ahrling2019transport, kyrtsos2018, zachinskis2023, green2022, tsao2018, yuan2021}. In contrast, Pd substitution induces a p-type behaviour  in the Ga\textsubscript{2}O\textsubscript{2} ML, whereas the remaining TM substituents examined in this work preserve the n-type behavior.\\
%Evaluation of the conductivity ratio $\chi$ for the favorably adsorbed molecules reveals a strongly analyte-dependent electronic response. 

\noindent NH\textsubscript{3} adsorption does not significantly alter the position of the adsorption-induced states relative to the Pd-Ga\textsubscript{2}O\textsubscript{2} ML. The partially filled state remains at the Fermi level, located approximately $0.35eV$ above the VBE, closely resembling the Pd-Ga\textsubscript{2}O\textsubscript{2} ML (see Fig.~S6 in SI). As a result, the dominant excitation pathway remains between the VBE and this partially filled state, thereby retaining the predominantly hole-mediated (p-type) conductivity of the system. Consequently, the hole excitation energy remains essentially unchanged, suggesting that the conductivity of the system will remain largely unaffected by NH\textsubscript{3} adsorption. Accordingly, NH\textsubscript{3} is not identified as a target analyte for Pd-Ga\textsubscript{2}O\textsubscript{2} ML.\\

\noindent In contrast, for H\textsubscript{2}S, H\textsubscript{2}O, CS\textsubscript{2}, and SO\textsubscript{2} adsorption, the dominant excitation occurs in CB. In the case of H\textsubscript{2}S and H\textsubscript{2}O, the electron excitation energy ($0.32eV$ and $0.37eV$, respectively) to CBE is comparable to the hole excitation energy to the VBE in the Pd-substituted ML. As a result, the net carrier excitation energy remains largely unchanged. Under these circumstances, the lower electron mobility becomes the dominant distinguishing factor, suggesting a reduction in conductivity upon the adsorption of these species. Nevertheless, owing to the comparable excitation energies associated with the electron mediated transport channel in the analyte adsorbed Pd-Ga\textsubscript{2}O\textsubscript{2} ML and hole mediated transport channels in the Pd-Ga\textsubscript{2}O\textsubscript{2} ML without the analyte, the magnitude of the resulting conductivity modulation cannot be inferred reliably from the present analysis since the two transport channels are associated with different carrier mobility. However we do not expect a reliable significant change in the conductivity  and the change may be modulated by other factors including unintentional impurity or defects in the ML.\\ %In both cases, the adsorption-induced redistribution of electronic states near the Fermi level is expected to perturb the baseline transport characteristics of the Pd-substituted ML, thereby contributing to the sensing response.\\

\noindent In the case of CS\textsubscript{2} and SO\textsubscript{2} adsorption, we note electron excitation energy of $0.27eV$ and $0.28eV$ respectively, which are smaller than the corresponding hole excitation energy. This might result in enhanced carrier excitations in the former case. However, it is also noteworthy that the electron mobility ($\mu_{e}$) in Ga\textsubscript{2}O\textsubscript{2} is an order of magnitude less compared to the hole mobility ($\mu_{h}$)~\cite{shao2021two}, which compensates the possibility of conductivity enhancement due to stronger electron excitation in the molecule adsorbed ML to some extent. These competing factors make it difficult to reliably predict either the magnitude or the polarity of the resulting conductivity modulation from the present analysis alone. Accordingly, the precise conductivity response of the all the four aforementioned molecules such as H\textsubscript{2}S, H\textsubscript{2}O, CS\textsubscript{2}, and SO\textsubscript{2} adsorbed systems remains beyond the scope of the present theoretical study and is left for future experimental investigation. However, we do not expect sufficient conductometric response from adsoprtion of these molecules and hence these are not considered as target molecules for detection by Pd-Ga\textsubscript{2}O\textsubscript{2} ML.\\
%This reduced excitation gap would, in isolation, be expected to enhance the conductivity. However, the dominant transport channel simultaneously shifts from hole-mediated to electron-mediated conduction. Since the electron mobility ($\mu_{e}$) is approximately an order of magnitude lower than the hole mobility ($\mu_{h}$), these two effects act in opposition. Consequently, the net change in conductivity cannot be unambiguously inferred from the present analysis and is left for future experimental investigation.\\

\noindent In the case of CO\textsubscript{2} adsorption, the dominant excitation occurs in the CB. In this case, the electron excitation energy of $0.53eV$ is larger than the hole excitation energy of the Pd-Ga\textsubscript{2}O\textsubscript{2} ML (see Fig. \ref{fig:DOS_Pd}). This is expected to suppress the conductivity. Furthermore, since the dominant transport channel is electron-mediated in the adsorbed ML and the electron mobility ($\mu_{e}$) is substantially lower than the hole mobility ($\mu_{h}$), this effect is anticipated to further reinforce the conductivity suppression. Consequently, CO\textsubscript{2} may serve as a viable target analyte for reusable sensing on the Pd-Ga\textsubscript{2}O\textsubscript{2} ML. Although CO\textsubscript{2} is an ambient species, such a response could be particularly useful in controlled or enclosed environments where accurate monitoring of CO\textsubscript{2} concentration is of practical interest. \\

\noindent NO\textsubscript{2} adsorption also substantially alters the electronic structure of the Pd-Ga\textsubscript{2}O\textsubscript{2} ML. In this case, we note a hole excitation energy (to the VB) of $0.95eV$ and electron excitation energy (to the CB) of $0.84eV$.
Consequently, the excitation of electrons and holes  are associated with significantly larger energy requirement compared to that of Pd-Ga\textsubscript{2}O\textsubscript{2} ML, leading to a suppression of the electrical conductivity.\\

\noindent It is noted that, unlike the aforementioned molecules, NO exhibits an adsorption energy of $-1.20eV$, placing it in the borderline adsorption regime where reusable sensing remains feasible, although complete surface recovery may require thermal annealing or UV-assisted desorption. DOS analysis reveals that NO adsorption eliminates the partially filled Fermi-level state present in the Pd-Ga\textsubscript{2}O\textsubscript{2} ML. Consequently, the relevant excitations are governed by the gap between VBE and the lowest unoccupied state, located approximately at $1.2eV$ above it or on the CB and the highest occupied state located at $0.88eV$ below it. This substantial increase in the effective excitation energy relative to the  Pd-Ga\textsubscript{2}O\textsubscript{2} ML is expected to significantly suppress the electrical conductivity. Thus, NO, NO\textsubscript{2} and CO\textsubscript{2} are identified as suitable target analytes for reusable sensing on the Pd-Ga\textsubscript{2}O\textsubscript{2} ML, exhibiting both favorable adsorption characteristics and adsorption-induced conductivity suppression. \\

\noindent \textit{\textbf{Selectivity Analysis of NO and NO\textsubscript{2} sensing under ambient conditions}}: To understand the selectivity of the Pd-Ga\textsubscript{2}O\textsubscript{2} ML in the ambient environment, we need to understand the detectability of these gases in the presence of CO\textsubscript{2}. As discussed above, in addition to NO and NO\textsubscript{2}; CO\textsubscript{2} also exhibits stable adsorption on Pd-Ga\textsubscript{2}O\textsubscript{2} and is predicted to induce a measurable conductometric response. Since CO\textsubscript{2} is present at reasonable concentrations, it is important to evaluate whether competition for active sites from CO\textsubscript{2}, hampers the detectivity of NO and NO\textsubscript{2} in the ambient environment. To this end, adsorption site occupancy probabilities were computed using the grand canonical adsorption model (Section~3.1, Eq.~S3, SI)~\cite{chen2024transition} considering NO and NO\textsubscript{2} at 25 ppm  (regulatory permissible exposure limit~\cite{osha_pel, NIOSH_NO}) in the presence of O\textsubscript{2} (21\%), alongside N\textsubscript{2} ($79\%$), H\textsubscript{2}O (5000 ppm), and CO\textsubscript{2} (400 ppm) at 1 atm total pressure. Detailed active site occupancy probability calculations are provided in Section~2.1 of the SI.\\

\noindent For NO, the exceptionally strong adsorption energy ($-1.20eV$) overwhelmingly outweighs the concentration advantage of the ambient gases, yielding a site occupancy probability approaching unity (see Table S1, SI). Consequently, NO emerges as the dominant surface species under ambient conditions, with negligible competition from O\textsubscript{2}, H\textsubscript{2}O, N\textsubscript{2} or CO\textsubscript{2}. This observation further reinforces the suitability of Pd-Ga\textsubscript{2}O\textsubscript{2} for practical NO detection.\\

\noindent In contrast, the occupancy probability of NO\textsubscript{2} is found to be only $\sim0.26\%$ with CO\textsubscript{2} having comparable probability of $\sim-0.20\%$ as well, under identical conditions (see Table S2, SI). The active sites are predominantly occupied by H\textsubscript{2}O ($\sim97.98\%$), while there is negligible probability for O\textsubscript{2} and N\textsubscript{2} occupancy. Therefore, despite its favorable adsorption energy and predicted conductometric response, practical NO\textsubscript{2} detection is expected to be significantly hindered under ambient conditions due to competition from humidity, and to a lesser extent CO\textsubscript{2}. These results suggest that NO\textsubscript{2} sensing using Pd-Ga\textsubscript{2}O\textsubscript{2} would be more feasible in humidity and CO\textsubscript{2}-controlled environments. Thus, NO emerges as a promising target analyte under ambient conditions; however, its relatively strong adsorption suggests that reusable operation would likely require assisted recovery mechanisms.\\

 %Unlike NH\textsubscript{3} adsorption, the conductivity response in this case is governed primarily by transitions involving states on the conduction-band side of the electronic structure, thereby imparting a predominantly electron-mediated (n-type-like) transport character. Since the electron mobility ($\mu_{e}$) in Ga\textsubscript{2}O\textsubscript{2} is approximately one order of magnitude lower than the hole mobility ($\mu_{h}$)~\cite{shao2021two}, the shift toward an electron-mediated transport channel may further contribute to the suppression of conductivity.\\

%It is noted that NO, with an adsorption energy of $-1.20eV$, sits in the borderline strong chemisorption regime, where complete surface regeneration under ambient conditions is impractical and would necessitate external thermal or UV treatment. %The Pd–Ga\textsubscript{2}O\textsubscript{2} ML exhibits a strongly analyte-dependent sensing response. 
%\noindent It should be noted that in Pd-Ga\textsubscript{2}O\textsubscript{2}, adsorption of some gas molecules switches the nature of the overall adsorbed system to n-type. In particular, adsorption of NH\textsubscript{3}, results in lower hole excitation energy compared to electron excitation energy and preserves the the p-type nature. Adsorption of CS\textsubscript{2}, SO\textsubscript{2}, NO, NO\textsubscript{2}, and CO\textsubscript{2} on the other hand results in lower electron excitation energy compared to hole excitation energy and switches the overall nature from p-type to n-type. 

\subsection{Zn, Zr and Mo Substitution}
\label{subsec:Zn_sub}
\textit{\textbf{Zn-substituted System}}: Among the substituted configurations examined, the Zn-substituted ML (Zn-Ga\textsubscript{2}O\textsubscript{2}) presents a particularly nuanced functional picture. The adsorption energies for molecules such as CO\textsubscript{2} ($E_{ads}$=$-0.25eV$), CS\textsubscript{2} ($-0.35eV$), and HF ($-0.34eV$) fall weaker than $-0.4eV$ threshold, showing unstable adsorption, with recovery times in the sub-microsecond range, precluding any stable signal detection. Since these molecules exhibit weak adsorption characteristics, a detailed DOS analysis was not performed for them.\\ 

\begin{figure}[htbp]
	\centering
	\includegraphics[width=0.99\columnwidth]{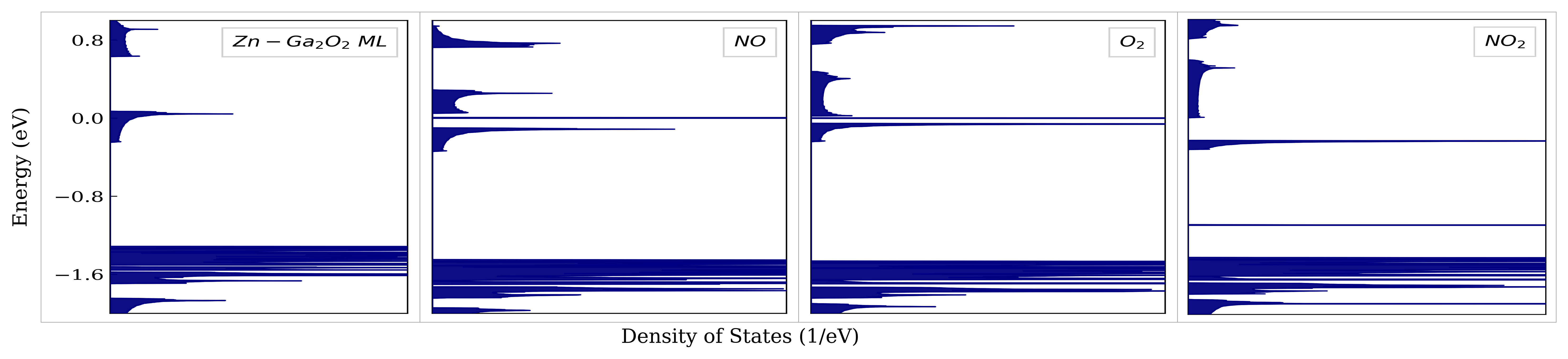}
	\caption{Total DOS of the Zn-Ga\textsubscript{2}O\textsubscript{2} ML and upon adsorption of  the favorable target molecules. The Fermi level is pinned at $0eV$ in all the panels.}
	\label{fig:Zn_DOS}
\end{figure}

\noindent Molecules such as NO ($-0.62eV$), SO\textsubscript{2} ($-0.54eV$), CO ($-0.49eV$), H\textsubscript{2}S ($-0.80eV$), H\textsubscript{2}O ($-0.78eV$), and O\textsubscript{2} ($-0.95eV$) exhibit adsorption energies within the range between $-0.4eV$ to $-1.0eV$ for reusable gas sensors. DOS analysis of the Zn-Ga\textsubscript{2}O\textsubscript{2} ML (see Fig.~\ref{fig:Zn_DOS}) reveals a delocalised mini-band like characteristics at the Fermi level, endowing the surface with intrinsic metallic character~\cite{Davies1998, capasso1986, Sibille1990, esaki1970}. For adsorbed SO\textsubscript{2}, CO, H\textsubscript{2}O, and H\textsubscript{2}S, the Fermi level remains within this mini-band-like state (see Fig.~S9 in SI), indicating that the conductivity of the adsorbed systems remains comparable to that of the Zn-substituted monolayer. Therefore, these molecules are not considered as the target analytes for Zn-Ga\textsubscript{2}O\textsubscript{2} ML. However, upon adsorption of NO and O\textsubscript{2}, the Fermi energy moves outside this delocalised miniband into the forbidden gap, giving rise to an electron excitation energy of $0.15eV$ ($\approx$6$k_{B}T$) and $0.075eV$ ($\approx$3$k_{B}T$), respectively at $T=300K$ (see Fig.~\ref{fig:Zn_DOS}). This indicates a metal-to-semiconductor transition accompanied by a significant reduction in electrical conductivity.\\ %Since the adsorption is marked by a change in nature from metallic to semiconducting, a reliable quantitative estimate of the conductivity modulation cannot be obtained within the present framework and is therefore left for future experimental studies.\\
% Nevertheless, these gaps remain small and are readily bridged by thermal fluctuations at room temperature.\\

\begin{figure*}[htbp]
	\centering
	
	\begin{minipage}[c]{0.50\textwidth}
		\centering
		\scriptsize
		\setlength{\tabcolsep}{6pt}
		\renewcommand{\arraystretch}{1.3}
		
		\begin{tabular}{lcccc}
			\hline
			Molecule & $E_{ads}$ (eV) & H (\AA) & q (e) & $\tau$ (s) \\
			\hline
			NH\textsubscript{3}  & -1.30 & 2.08 & -0.16 & $6.90 \times 10^{9}$ \\
			NO                  & -0.62 & 2.11 & 0.13 & $2.60 \times 10^{-2}$ \\
			SO\textsubscript{2} & -0.54 & 1.99 & 0.19 & $1.18 \times 10^{-3}$ \\
			O\textsubscript{2}  & -0.95 & 1.97 & 0.42 & $9.11 \times 10^{3}$ \\
			NO\textsubscript{2} & -1.37 & 2.01 & 0.51 & $1.04 \times 10^{11}$ \\
			H\textsubscript{2}S & -0.80 & 2.45 & -0.16 & $2.75 \times 10^{1}$ \\
			H\textsubscript{2}O & -0.78 & 2.13 & -0.05 & $1.27 \times 10^{1}$ \\
			CO                  & -0.49 & 2.15 & -0.05 & $1.70 \times 10^{-4}$ \\
			CO\textsubscript{2} & -0.25 & 2.37 & -0.0008 & $1.58 \times 10^{-8}$ \\
			CS\textsubscript{2} & -0.35 & 2.97 & -0.01 & $7.58 \times 10^{-7}$ \\
			HF                  & -0.34 & 2.27 & -0.01 & $5.15 \times 10^{-7}$ \\
			\hline
		\end{tabular}
		
		\vspace{2.0em}
		{\normalsize\textbf{(a)}}
	\end{minipage}
	\hfill
	\begin{minipage}[c]{0.48\textwidth}
		\centering
		\includegraphics[width=0.95\linewidth]{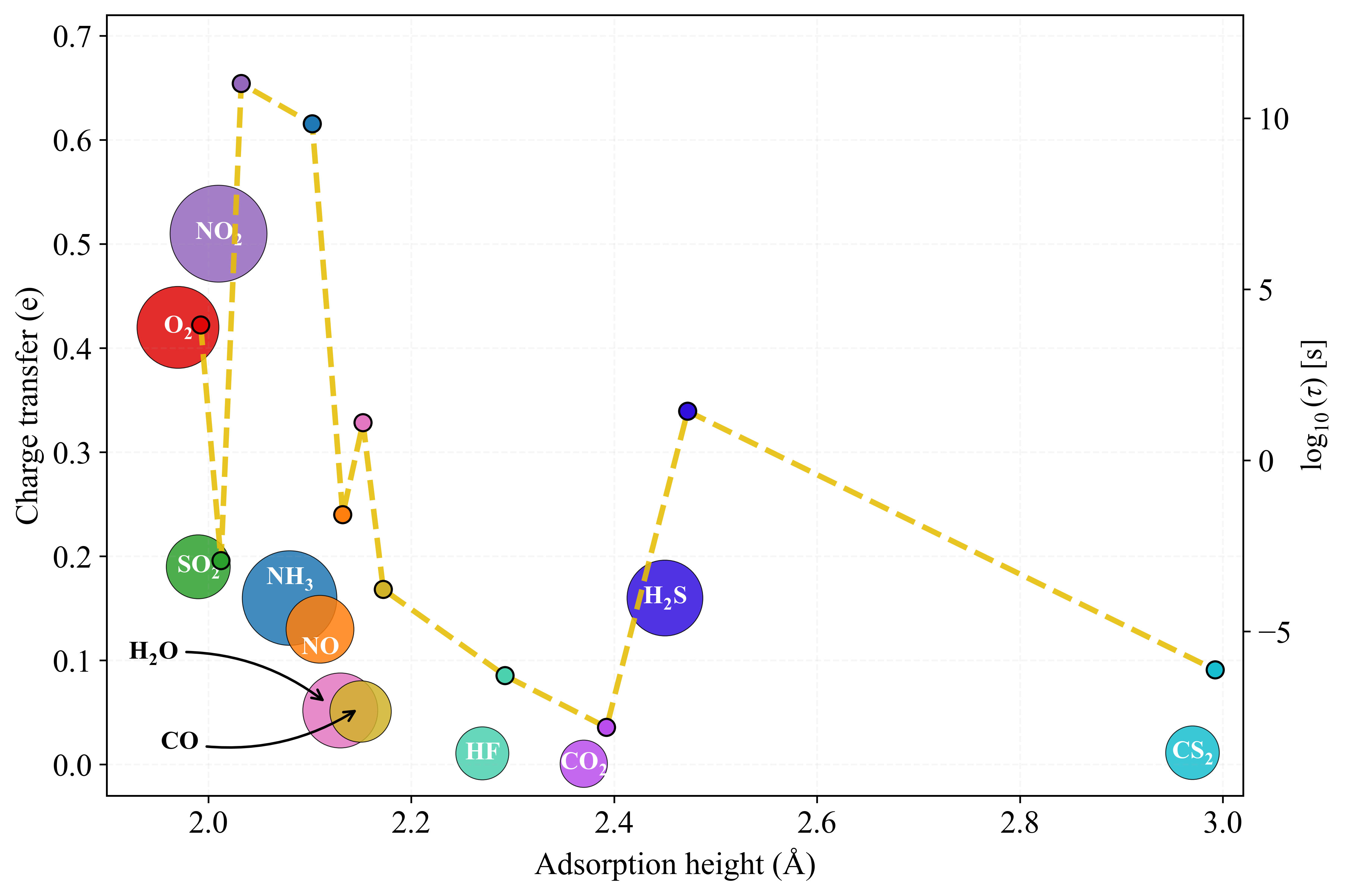}
		
		\vspace{0.3em}
		\textbf{(b)}
	\end{minipage}
	
	\caption{(a) Table II: Adsorption parameters of the investigated gas molecules on the Zn-Ga\textsubscript{2}O\textsubscript{2} ML, including adsorption energy ($E_{ads}$), adsorption height (H), charge transfer (q), and recovery time ($\tau$). (b) Graphical summary of the adsorption characteristics of gas molecules on the Zn-Ga\textsubscript{2}O\textsubscript{2} monolayer. The centers of the larger bubbles are referenced to the X and primary Y-axes, representing the adsorption height and charge transfer, respectively. The area of each bubble is proportional to the magnitude of the adsorption energy. The smaller markers linked by the dashed line indicate the corresponding recovery times, referenced to the secondary Y-axis. Recovery times are reported on a logarithmic scale.}
	\label{fig:Zn_summary}
\end{figure*}

\noindent In contrast to the above case, NH\textsubscript{3} ($-1.30eV$) and NO\textsubscript{2} ($-1.37eV$) exhibit strong chemisorptive interactions approaching the $-1.5eV$ irreversibility threshold. Their low adsorption heights and substantial charge transfer (see Fig.~\ref{fig:Zn_summary}) further point towards deep chemisorptive adsorption, while the predicted recovery times (see Fig.~\ref{fig:Zn_summary}) suggest that surface recovery may only be achievable through external thermal annealing or UV-assisted desorption techniques. Despite the strong adsorption of NH\textsubscript{3}, it does not significantly alter the electronic structure of the Zn-Ga\textsubscript{2}O\textsubscript{2} ML (see Fig.~S9 in SI). The Fermi level continues to penetrate the miniband as present in the Zn-substituted ML, thereby preserving its metallic character. Consequently, NH\textsubscript{3} is not considered a target analyte for the Zn-Ga\textsubscript{2}O\textsubscript{2} ML. On the contrary, the adsorption of NO\textsubscript{2} opens a considerably larger energy gap of about $0.235eV$ ($\approx$10$k_{B}T$ at $T=300K$), indicating substantial conductivity suppression and a transition from metallic to semiconducting behavior. %Owing to this transition, quantitative estimation of the conductivity change in this case cannot be reliably obtained within the present framework. 
Nevertheless, as the adsorption energy of NO\textsubscript{2} remains below the $-1.5~\mathrm{eV}$ irreversibility criterion, Zn-Ga\textsubscript{2}O\textsubscript{2} may still be considered a potential reusable resistive sensing platform for NO\textsubscript{2}, provided that externally assisted recovery techniques are employed.\\

\noindent \textit{\textbf{Selectivity Analysis of NO and NO\textsubscript{2} sensing under ambient conditions:}} It is worth noting that O\textsubscript{2} is an ambient species, but also emerges as a favorable target analyte on Zn-Ga\textsubscript{2}O\textsubscript{2}, exhibiting both a suitable adsorption energy ($-0.95eV$) and a pronounced conductometric response. To assess the practical viability of NO and NO\textsubscript{2} detection under ambient conditions, adsorption site occupancy probabilities were evaluated using the same method utilised for Pd-Ga\textsubscript{2}O\textsubscript{2}. The detailed calculations are provided in the Section~4.1 in SI.\\

\noindent For NO, the calculated occupancy probability is quite negligible ($\sim10^{-5}$), while O\textsubscript{2} occupies nearly all available active sites $\sim99.99\%$. This indicates that competitive adsorption from atmospheric O\textsubscript{2} would severely hinder practical NO detection under ambient conditions. In contrast, NO\textsubscript{2} exhibits an occupancy probability of $\sim99.87\%$, substantially exceeding that of O\textsubscript{2} $\sim0.13\%$. Therefore, despite its much lower concentration, NO\textsubscript{2} effectively outcompetes atmospheric oxygen owing to its significantly stronger adsorption. Consequently, while NO sensing is expected to require O\textsubscript{2}-controlled environments, NO\textsubscript{2} detection should remain viable under ambient conditions, albeit with the requirement of externally assisted recovery owing to its strong chemisorptive interaction.\\

\begin{figure*}[htbp]
	\centering
	
	\begin{minipage}[c]{0.50\textwidth}
		\centering
		\scriptsize
		\setlength{\tabcolsep}{6pt}
		\renewcommand{\arraystretch}{1.3}
		
		\begin{tabular}{lcccc}
			\hline
			Molecule & $E_{ads}$ (eV) & H (\AA) & q (e) & $\tau$ (s) \\
			\hline
			NH\textsubscript{3}  & -1.39 & 2.42 & -0.097 & $2.24 \times 10^{11}$ \\
			NO                  & -1.42 & 2.06 & 0.435 & $7.16 \times 10^{11}$ \\
			SO\textsubscript{2} & -0.37 & 2.71 & 0.404 & $1.64 \times 10^{-6}$ \\
			O\textsubscript{2}  & -2.29 & 1.92 & 0.6858 & $2.95 \times 10^{26}$ \\
			NO\textsubscript{2} & -2.86 & 1.96 & 0.6274 & $1.11 \times 10^{36}$ \\
			H\textsubscript{2}S & -0.81 & 2.90 & -0.0757 & $4.05 \times 10^{1}$ \\
			H\textsubscript{2}O & -0.82 & 2.47 & -0.0265 & $5.96 \times 10^{1}$ \\
			CO                  & -0.52 & 2.60 & 0.0149 & $5.44 \times 10^{-4}$ \\
			CO\textsubscript{2} & -0.09 & 3.45 & 0.0265 & $3.25 \times 10^{-11}$ \\
			CS\textsubscript{2} & -0.51 & 2.79 & 0.0054 & $3.70 \times 10^{-4}$ \\
			HF                  & -0.53 & 2.46 & -0.004 & $8.01 \times 10^{-4}$ \\
			\hline
		\end{tabular}
		
		\vspace{2.0em}
		{\normalsize\textbf{(a)}}
	\end{minipage}
	\hfill
	\begin{minipage}[c]{0.48\textwidth}
		\centering
		\includegraphics[width=0.95\linewidth]{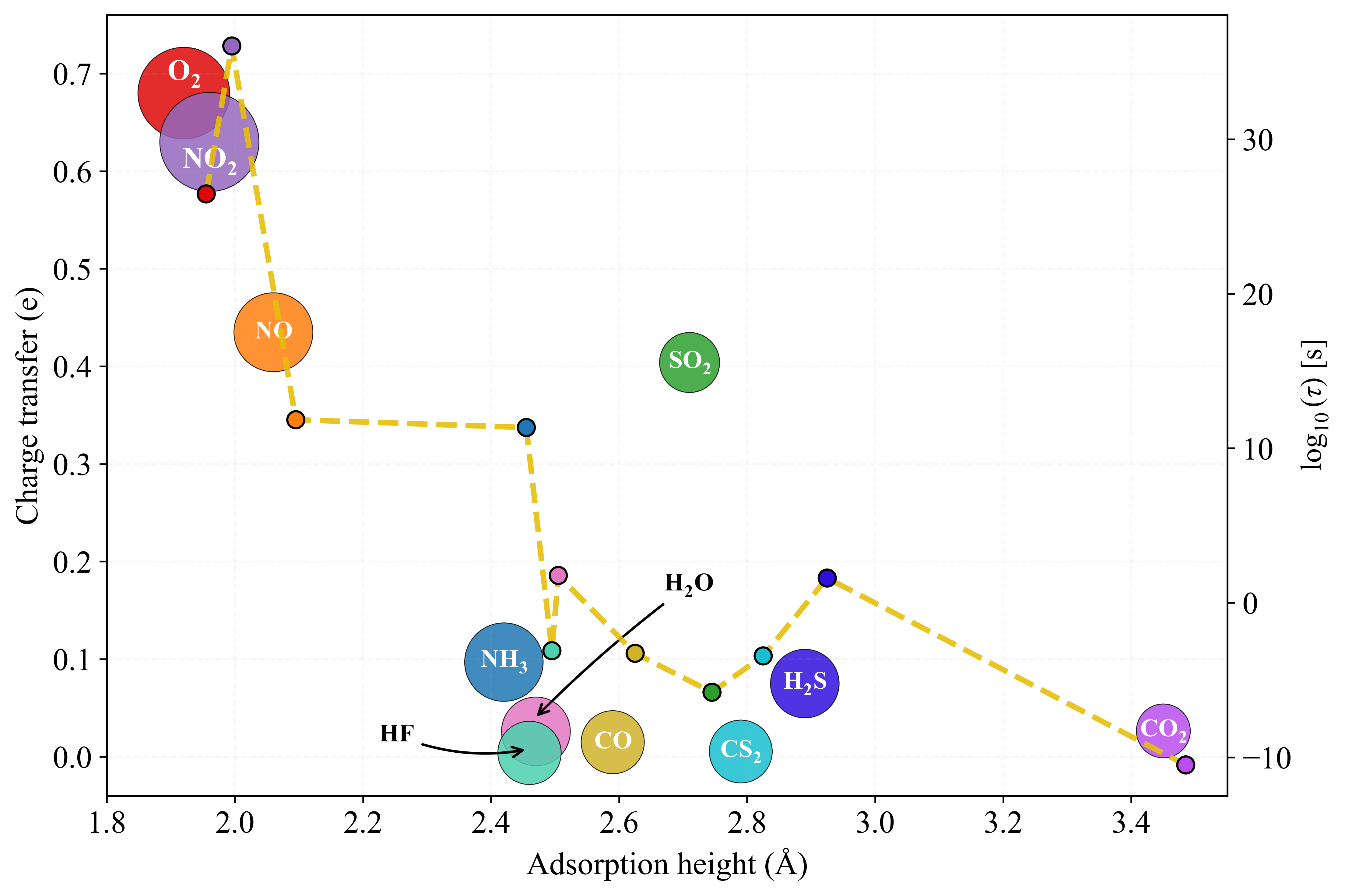}
		
		\vspace{0.3em}
		\textbf{(b)}
	\end{minipage}
	
	\caption{(a) Table III: Adsorption parameters of the investigated gas molecules on the Zr-Ga\textsubscript{2}O\textsubscript{2} ML, including adsorption energy ($E_{ads}$), adsorption height (H), charge transfer (q), and recovery time ($\tau$). (b) Graphical summary of the adsorption characteristics of gas molecules on the Zr-Ga\textsubscript{2}O\textsubscript{2} monolayer. The centers of the larger bubbles are referenced to the X and primary Y-axes, representing the adsorption height and charge transfer, respectively. The area of each bubble is proportional to the magnitude of the adsorption energy. The smaller markers linked by the dashed line indicate the corresponding recovery times, referenced to the secondary Y-axis. Recovery times are reported on a logarithmic scale.}
	\label{fig:Zr_summary}
\end{figure*}

\begin{figure*}[htbp]
	\centering
	\begin{subfigure}{0.48\textwidth}
		\centering
		\includegraphics[width=\linewidth]{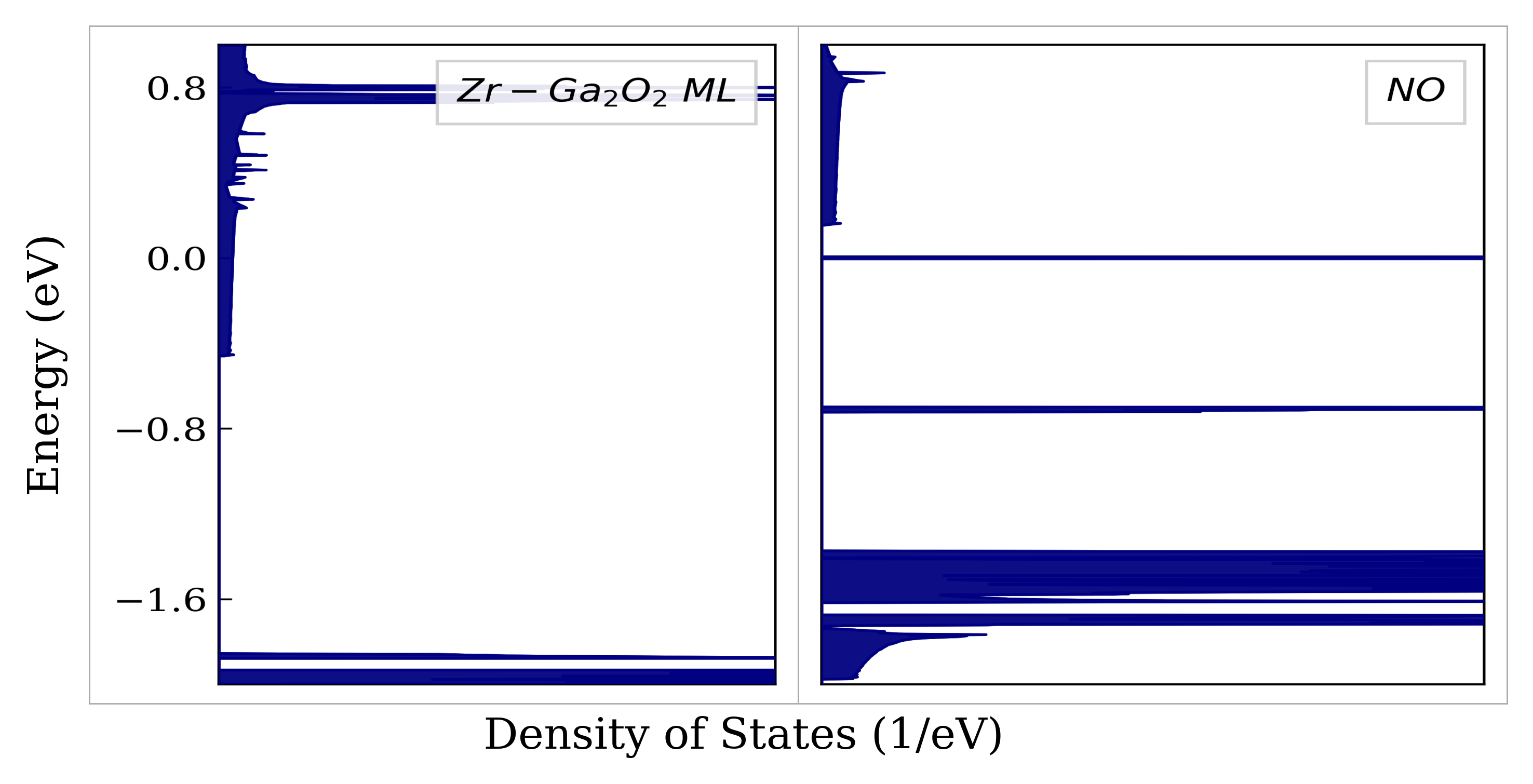}
		\caption{}
		\label{fig:Zr_target}
	\end{subfigure}
	\hfill
	\begin{subfigure}{0.48\textwidth}
		\centering
		\includegraphics[width=\linewidth]{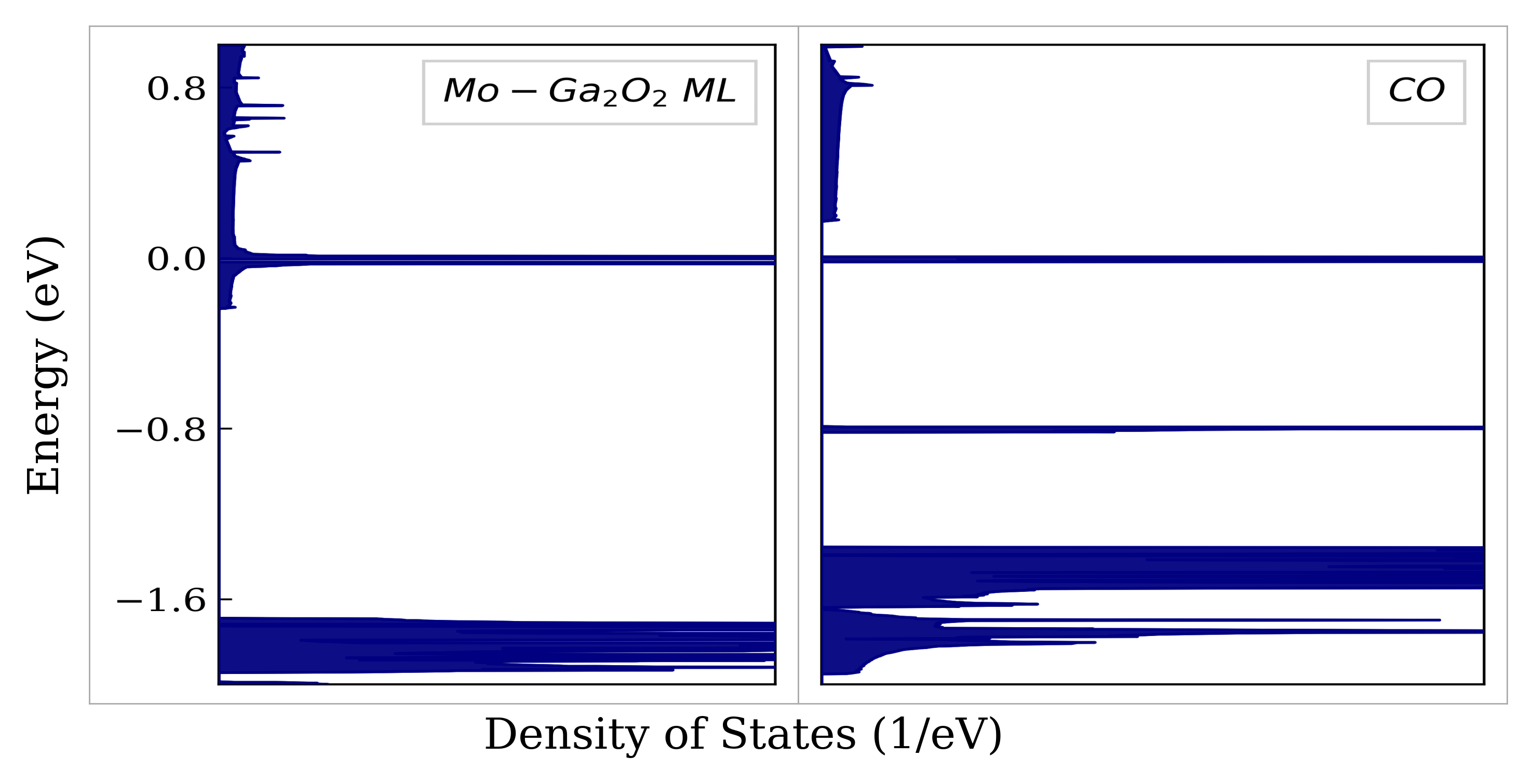}
		\caption{}
		\label{fig:Mo_target}
	\end{subfigure}
	\caption{Total DOS of the TM-Ga\textsubscript{2}O\textsubscript{2} ML and upon adsorption of the favorable gas molecules: (a) Zr-Ga\textsubscript{2}O\textsubscript{2} ML and its target analyte. (b): Mo-Ga\textsubscript{2}O\textsubscript{2} ML and its target analyte. The Fermi level is pinned at $0eV$ in all the panels.}
	\label{fig:Mo_Zr_DOS}
\end{figure*}

\noindent \textit{\textbf{Zr-substituted System}}: In the Zr-substituted ML (Zr-Ga\textsubscript{2}O\textsubscript{2}), molecules such as SO\textsubscript{2} ($E_{ads}$=$-0.37eV$) and CO\textsubscript{2} ($-0.09eV$) show weak adsorption energies with $E_{ads}$ $>$ $-0.4eV$. The DOS were not analysed in these cases. On the other hand, a subset of analytes like H\textsubscript{2}S ($-0.81eV$), H\textsubscript{2}O ($-0.82eV$), CO ($-0.52eV$), CS\textsubscript{2} ($-0.51eV$), and HF ($-0.53eV$) exhibit adsorption energies within the $-0.4eV$ to $-1.0eV$ window, satisfying the energy criterion for stable and reversible adsorption. The values of the adsorption energies for all the molecules under consideration have been summarized in Fig.~\ref{fig:Zr_summary}. DOS analysis of the Zr-Ga\textsubscript{2}O\textsubscript{2} ML reveals that the Fermi energy penetrates the conduction band, conferring intrinsic metallic character (see Fig.~\ref{fig:Zr_target}). For these five molecules with $E_{ads}$ in the range of $-0.4eV$ to $-1eV$, the Fermi energy of the Zr-Ga\textsubscript{2}O\textsubscript{2} ML  continues to penetrate the CB upon adsorption, demonstrating metallic behaviour like that of the Zr-Ga\textsubscript{2}O\textsubscript{2} ML. Hence, the conductometric change factor may not be strong enough for robust and reliable detection. In the absence of a strong conductometric signal change, these molecules, despite their favorable adsorption energetics are rendered electrically invisible. Consequently, they cannot be classified as suitable targets for reusable resistive gas sensing on this platform.\\
%\begin{figure}[htbp]
%	\centering
%	\includegraphics[width=0.8\columnwidth]{Pictures/Zr_Mo_target_collage}
%	\caption{Total DOS of the TM-Ga\textsubscript{2}O\textsubscript{2} ML and upon adsorption of the favorable gas molecules:(Left) Zr-Ga\textsubscript{2}O\textsubscript{2} ML and its target analytes. (Right): Mo-Ga\textsubscript{2}O\textsubscript{2} ML and its target analytes. The Fermi level is pinned at $0eV$ in all the panels.}
%	\label{fig:Mo_Zr_DOS}
%\end{figure}

\noindent In contrast, NH\textsubscript{3} ($-1.39eV$) and NO ($-1.42eV$) exhibit strong chemisorptive adsorption behaviour approaching but not exceeding the $-1.5eV$ irreversibility threshold. Surface recovery, while impractical under ambient conditions, may in principle be achieved through external thermal annealing or UV photo-desorption treatment. The system remains metallic on adsorption of NH\textsubscript{3}, indicating that the conductivity will not change by an appreciable factor. Consequently, NH\textsubscript{3} is unsuitable as a target analyte for resistive gas sensing. On the other hand, the adsorption of NO shifts the Fermi level out of the CB, thereby inducing an electron excitation energy of $0.85eV$ ($\approx$33$k_{B}T$ at $T=300K$) to the CB (see Fig.~\ref{fig:Zr_target}). This indicates a transition from metallic to semiconducting behavior, leading to a substantial reduction in conductivity. Therefore, NO can be identified as a promising target analyte for reusable resistive gas sensing with assisted recovery  but in O\textsubscript{2} restricted environment (see below).\\%It should be noted that a quantitative estimate of conductivity suppression upon NO adsorption lies beyond the scope of the present study due to metallic to semiconducting transition upon adsorption.
\noindent O\textsubscript{2} ($-2.29eV$) and NO\textsubscript{2} ($-2.86eV$) far exceed the $-1.5eV$ threshold for molecular scavenging, exhibiting markedly reduced adsorption heights, substantial charge transfer, and recovery times extending from years to decades. For both molecules, adsorption drives the Fermi level out of the CB, transforming the metallic Zr-Ga\textsubscript{2}O\textsubscript{2} ML into a semiconducting system (see Fig.~S12 in SI). In the case of O\textsubscript{2}, an excitation energy of approximately $0.49eV$ is established from the partially occupied state to the CBE, while NO\textsubscript{2} induces a larger excitation energy of $0.68eV$. These adsorption-induced electronic modifications are expected to strongly suppress the conductivity of the system. However, owing to the exceptionally strong adsorption strength of O\textsubscript{2} and NO\textsubscript{2}, the interaction is more representative of irreversible capture than reusable sensing application. Collectively, these characteristics indicate that the Zr-Ga\textsubscript{2}O\textsubscript{2} ML exhibits potential for reusable resistive sensing of NO under controlled operating conditions where competitive adsorption from strongly adsorbed ambient species like O\textsubscript{2} is minimized. Conversely, the strong interactions with O\textsubscript{2} and NO\textsubscript{2} highlight the potential applicability of the Zr-Ga\textsubscript{2}O\textsubscript{2} ML for molecular scavenging of these gaseous species.\\

\noindent \textit{\textbf{Mo-substituted System}}: In the Mo-substituted ML (Mo-Ga\textsubscript{2}O\textsubscript{2}), the adsorption of SO\textsubscript{2} ($1.36eV$) is highly unstable in nature as its $E_{ads}$ is positive. Similarly, molecules such as CS\textsubscript{2} ($E_{ads}$=$-0.32eV$), HF ($-0.19eV$), and  CO\textsubscript{2} ($-0.07eV$) show weak adsorption behaviour with $E_{ads}$ $>$ $-0.4eV$. In these cases, the DOS has not been analysed in detail. On the other hand, H\textsubscript{2}S ($-0.61eV$), H\textsubscript{2}O ($-0.61eV$), and NH\textsubscript{3} ($-1.0eV$) exhibit adsorption energies within the $-0.4eV$ to $-1.0eV$ window, satisfying the energy criterion for stable and reversible adsorption.\\

%\begin{figure}[htbp]
%	\centering
%	\includegraphics[width=0.48\columnwidth]{Pictures/Zr_target_collage.png}
%	\hspace{0.02\columnwidth}
%	\includegraphics[width=0.48\columnwidth]{Pictures/Mo_target_collage.png}
%	\caption{}
%\end{figure}

\noindent DOS analysis of the Mo-Ga\textsubscript{2}O\textsubscript{2} ML reveals a similar behaviour to that of Zr-Ga\textsubscript{2}O\textsubscript{2} ML, wherein the Fermi level penetrates the conduction band, indicating intrinsic metallic character (see Fig.~\ref{fig:Mo_target}). Upon adsorption of H\textsubscript{2}S, H\textsubscript{2}O, and NH\textsubscript{3}, the system retains its metallic character, indicating that the conductivity change relative to the Mo-Ga\textsubscript{2}O\textsubscript{2} ML is not expected to be reliable enough for robust electrical detection (see Fig.~S15 in SI). As a result, these molecules are not considered as target analytes for Mo-Ga\textsubscript{2}O\textsubscript{2} ML.\\

\begin{figure*}[htbp]
	\centering
	
	\begin{minipage}[c]{0.50\textwidth}
		\centering
		\scriptsize
		\setlength{\tabcolsep}{6pt}
		\renewcommand{\arraystretch}{1.3}
		
		\begin{tabular}{lcccc}
			\hline
			Molecule & $E_{ads}$ (eV) & H (\AA) & q (e) & $\tau$ (s) \\
			\hline
		NH\textsubscript{3}  & -1.00 & 2.07 & -0.13 & $6.30 \times 10^{4}$ \\
		NO                  & -3.58 & 1.76 & 0.51 & $1.38 \times 10^{48}$ \\
		SO\textsubscript{2} & 1.36 & 2.28 & 0.34 & $1.42 \times 10^{-35}$ \\
		O\textsubscript{2}  & -2.77 & 1.80 & 0.49 & $3.42 \times 10^{34}$ \\
		NO\textsubscript{2} & -2.06 & 1.94 & 0.59 & $4.04 \times 10^{22}$ \\
		H\textsubscript{2}S & -0.60 & 2.46 & -0.09 & $1.20 \times 10^{-2}$ \\
		H\textsubscript{2}O & -0.61 & 1.99 & -0.05 & $1.77 \times 10^{-2}$ \\
		CO                  & -1.24 & 1.94 & 0.28 & $6.78 \times 10^{8}$ \\
		CO\textsubscript{2} & -0.07 & 3.07 & 0.02 & $1.50 \times 10^{-11}$ \\
		CS\textsubscript{2} & -0.32 & 3.22 & 0.005 & $2.38 \times 10^{-7}$ \\
		HF                  & -0.19 & 2.18 & -0.0005 & $1.56 \times 10^{-9}$ \\
			\hline
		\end{tabular}
		
		\vspace{2.0em}
		{\normalsize\textbf{(a)}}
	\end{minipage}
	\hfill
	\begin{minipage}[c]{0.48\textwidth}
		\centering
		\includegraphics[width=0.95\linewidth]{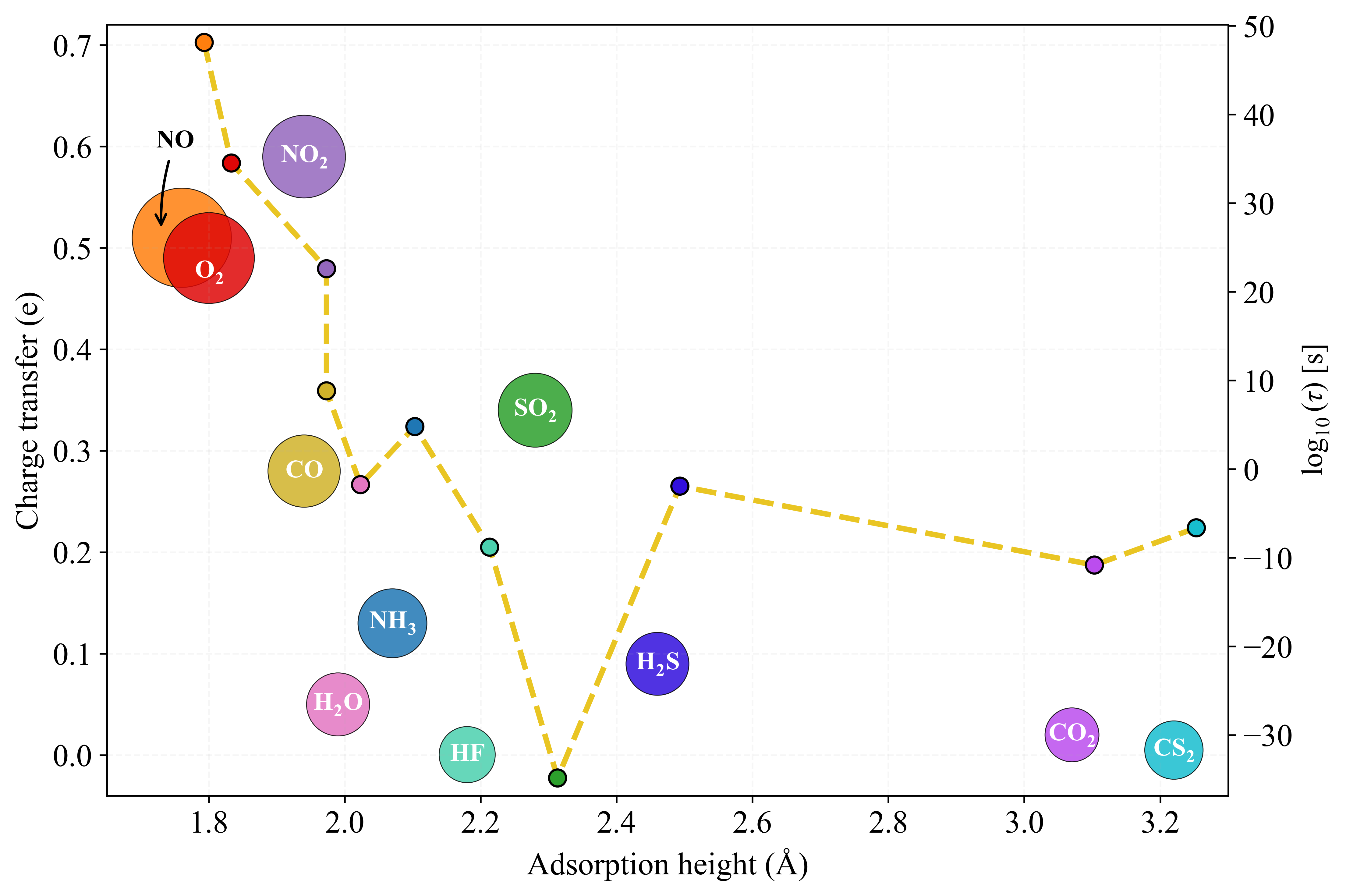}
		
		\vspace{0.3em}
		\textbf{(b)}
	\end{minipage}
	
	\caption{(a) Table IV: Adsorption parameters of the investigated gas molecules on the Mo-Ga\textsubscript{2}O\textsubscript{2} ML, including adsorption energy ($E_{ads}$), adsorption height (H), charge transfer (q), and recovery time ($\tau$). (b) Graphical summary of the adsorption characteristics of gas molecules on the Mo-Ga\textsubscript{2}O\textsubscript{2} monolayer. The centers of the larger bubbles are referenced to the X and primary Y-axes, representing the adsorption height and charge transfer, respectively. The area of each bubble is proportional to the magnitude of the adsorption energy. The smaller markers linked by the dashed line indicate the corresponding recovery times, referenced to the secondary Y-axis. Recovery times are reported on a logarithmic scale.}
	\label{fig:Mo_summary}
\end{figure*}

\noindent The adsorption of CO ($-1.24eV$) falls in the borderline regime beyond the reversible sensing threshold yet below the $-1.5eV$ irreversibility criterion. This indicates that the adsorption is too strong for practical surface recovery under ambient conditions, yet potentially amenable to external thermal annealing or UV photo-desorption treatment. CO adsorption drives the Fermi level out of the CB and induces a partially filled localized state near the CBE. The resulting excitation energy between this state and the CBE is approximately $0.164eV$ ($\approx$6$k_{B}T$) (Fig.~\ref{fig:Mo_target}). This is expected to reduce the electrical conductivity substantially, at room temperature. Therefore, the Mo-Ga\textsubscript{2}O\textsubscript{2} ML acts as a sensing layer to CO molecule, but in O\textsubscript{2} controlled environment (discussed below).\\

\noindent In sharp contrast, NO ($-3.58eV$), O\textsubscript{2} ($-2.77eV$), and NO\textsubscript{2} ($-2.06eV$) exhibit adsorption energies far exceeding the $-1.5eV$ threshold, accompanied by markedly low adsorption heights and substantial charge transfer. The corresponding values of $E_{ads}$, Q, H, and $\tau$ for all gas molecules on Mo-Ga\textsubscript{2}O\textsubscript{2} are summarized in Fig.~\ref{fig:Mo_summary}. The DOS remains largely unperturbed upon the adsorption of O\textsubscript{2}, thereby suggesting negligible modification of the electrical conductivity. On the contrary, the adsorption of NO and NO\textsubscript{2} shifts the Fermi level out of the energy bands, giving rise to electron excitation energy of $0.81eV$ and $0.012eV$, respectively (Fig.~S15 in the SI). While these changes are expected to significantly suppress the conductivity, the exceptionally strong adsorption energies of these molecules suggest that scavenging behavior would dominate over reusable sensing capability. Consequently, the Mo-Ga\textsubscript{2}O\textsubscript{2} ML appears more suitable for molecular scavenging applications toward NO, O\textsubscript{2}, and NO\textsubscript{2}. 

\subsection{Ag Substitution}
\label{subsec:Ag_sub}
\begin{figure*}[htbp]
	\centering
	
	\begin{minipage}[c]{0.50\textwidth}
		\centering
		\scriptsize
		\setlength{\tabcolsep}{6pt}
		\renewcommand{\arraystretch}{1.3}
		
		\begin{tabular}{lcccc}
			\hline
			Molecule & $E_{ads}$ (eV) & H (\AA) & q (e) & $\tau$ (s) \\
			\hline
			NH\textsubscript{3}  & -0.91  & 2.21  & 0.14  & $1.94 \times 10^{3}$ \\
			NO                   & -0.83 & 1.96 & 0.04 & $8.78 \times 10^{1}$ \\
			SO\textsubscript{2}  & -0.55 & 2.31 & 0.05 & $1.74 \times 10^{-3}$ \\
			O\textsubscript{2}  & -0.59 & 1.89 & 0.39 & $8.16 \times 10^{-3}$ \\
			NO\textsubscript{2}  & -0.75 & 2.05 & 0.22 & $3.98 \times 10^{0}$ \\
			H\textsubscript{2}S& -0.84 & 2.41 & 0.14 & $1.29 \times 10^{2}$ \\
			H\textsubscript{2}O& -0.45 & 2.38 & 0.05 & $3.63 \times 10^{-5}$ \\
			CO                 & -1.07 & 1.98 & 0.05 & $9.45 \times 10^{5}$ \\
			CO\textsubscript{2}& -0.16 & 2.59 & 0.009 & $4.87 \times 10^{-10}$ \\
			CS\textsubscript{2}& -0.65 & 2.21 & 0.04 & $8.31 \times 10^{-2}$ \\
			HF                 & -0.18 & 2.59 & 0.005 & $1.06 \times 10^{-9}$ \\
			\hline
		\end{tabular}
		
		\vspace{2.0em}
		{\normalsize\textbf{(a)}}
	\end{minipage}
	\hfill
	\begin{minipage}[c]{0.48\textwidth}
		\centering
		\includegraphics[width=0.95\linewidth]{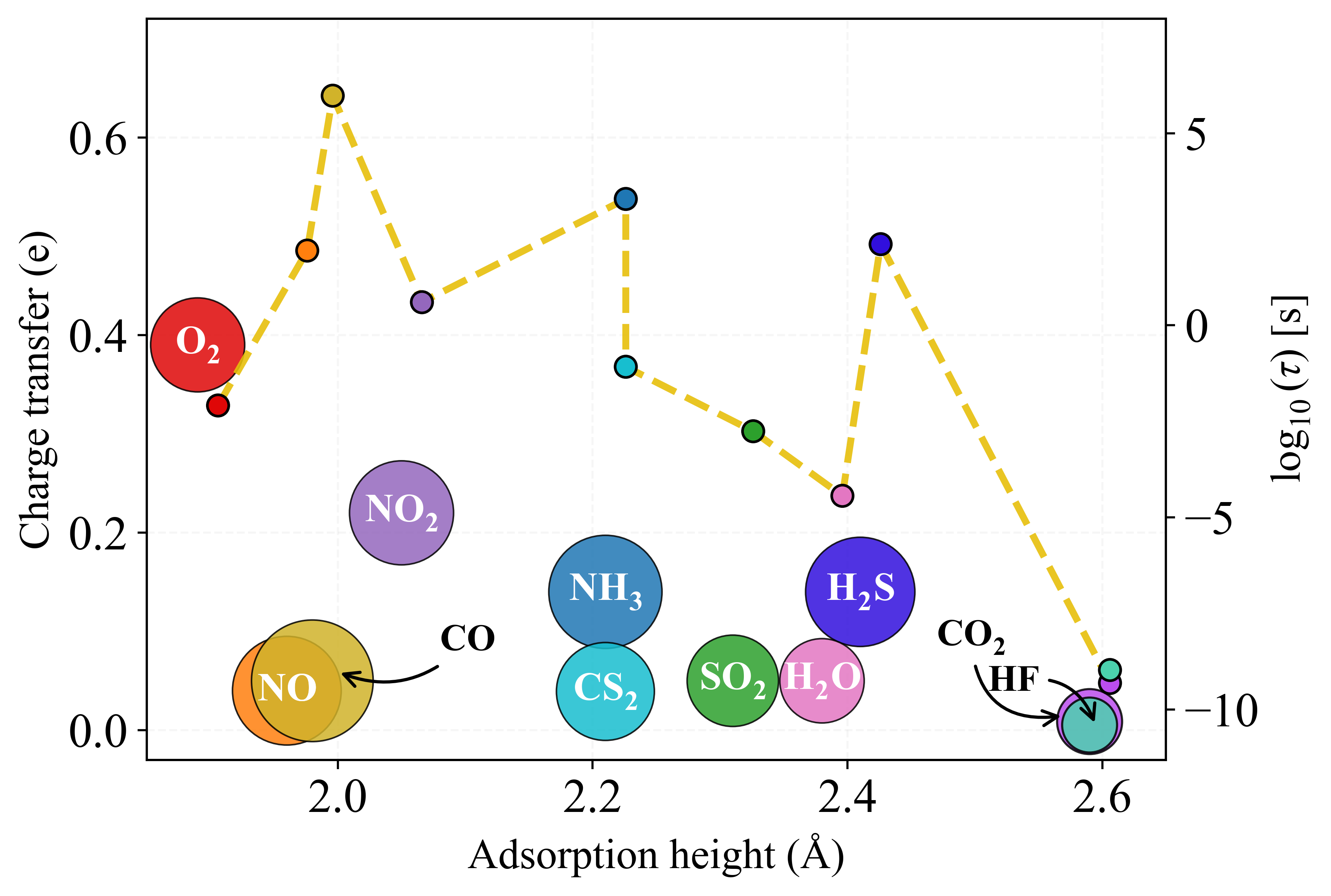}
		
		\vspace{0.3em}
		\textbf{(b)}
	\end{minipage}
	
	\caption{(a) Table V: Adsorption parameters of the investigated gas molecules on the Ag-Ga\textsubscript{2}O\textsubscript{2} ML, including adsorption energy ($E_{ads}$), adsorption height (H), charge transfer (q), and recovery time ($\tau$). (b) Graphical summary of the adsorption characteristics of gas molecules on the Ag-Ga\textsubscript{2}O\textsubscript{2} monolayer. The centers of the larger bubbles are referenced to the X and primary Y-axes, representing the adsorption height and charge transfer, respectively. The area of each bubble is proportional to the magnitude of the adsorption energy. The smaller markers linked by the dashed line indicate the corresponding recovery times, referenced to the secondary Y-axis. Recovery times are reported on a logarithmic scale.}
	\label{fig:Ag_summary}
\end{figure*}

%\begin{table}[ht]
%	\centering
%	\caption{Values of adsorption energy (E\textsubscript{ads}), charge transfer (q), recovery time ($\tau$), and adsorption height (H) for different molecules investigated on the Ag–Ga\textsubscript{2}O\textsubscript{2}}
%	\begin{tabular}{lcccc}
%		\hline
%		\textbf{Molecule} & \textbf{$E_{ads}$} (eV) & \textbf{H}(~\AA{}) & \textbf{q}(e) & \textbf{$\tau$}(s) \\
%		\hline
%		NH\textsubscript{3}  & -0.91  & 2.21  & 0.14  & $1.94 \times 10^{3}$ \\
%		NO                   & -0.83 & 1.96 & 0.04 & $8.78 \times 10^{1}$ \\
%		SO\textsubscript{2}  & -0.55 & 2.31 & 0.05 & $1.74 \times 10^{-3}$ \\
%		O\textsubscript{2}  & -0.59 & 1.89 & 0.39 & $8.16 \times 10^{-3}$ \\
%		NO\textsubscript{2}  & -0.75 & 2.05 & 0.22 & $3.98 \times 10^{0}$ \\
%		H\textsubscript{2}S& -0.84 & 2.41 & 0.14 & $1.29 \times 10^{2}$ \\
%		H\textsubscript{2}O& -0.45 & 2.38 & 0.05 & $3.63 \times 10^{-5}$ \\
%		CO                 & -1.07 & 1.98 & 0.05 & $9.45 \times 10^{5}$ \\
%		CO\textsubscript{2}& -0.16 & 2.59 & 0.009 & $4.87 \times 10^{-10}$ \\
%		CS\textsubscript{2}& -0.65 & 2.21 & 0.04 & $8.31 \times 10^{-2}$ \\
%		HF                 & -0.18 & 2.59 & 0.005 & $1.06 \times 10^{-9}$ \\
%		\hline
%	\end{tabular}
%	\label{tab:ag_adsorption_values}
%\end{table}
For the Ag-substituted ML (Ag-Ga\textsubscript{2}O\textsubscript{2}), based on the values of the adsorption energies, CO\textsubscript{2} ($E_{ads}$=$-0.16eV$) and HF ($-0.18eV$) are characterized by weak adsorption, as their adsorption energies lie above $-0.4eV$. Consequently, neither molecule satisfies the adsorption criterion adopted for practical gas sensing, and therefore their DOS are not analyzed further.\\

\noindent In contrast, molecules such as NH\textsubscript{3} ($E_{ads}$=$-0.91eV$), NO ($-0.83eV$), SO\textsubscript{2}($-0.55eV$), NO\textsubscript{2} ($-0.75eV$), H\textsubscript{2}S($-0.84eV$), CS\textsubscript{2}($-0.65eV$), CO($-1.07eV$) along with the ambient species H\textsubscript{2}O($-0.45eV$) and O\textsubscript{2} ($-0.59eV$), exhibit comparatively stronger adsorption with $E_{ads}$ lying within the favorable adsorption window. The values of $E_{ads}$, Q, H and $\tau$ are given in Fig.~\ref{fig:Ag_summary}.\\

\begin{figure*}[!htbp]
	\centering
	\includegraphics[width=\textwidth]{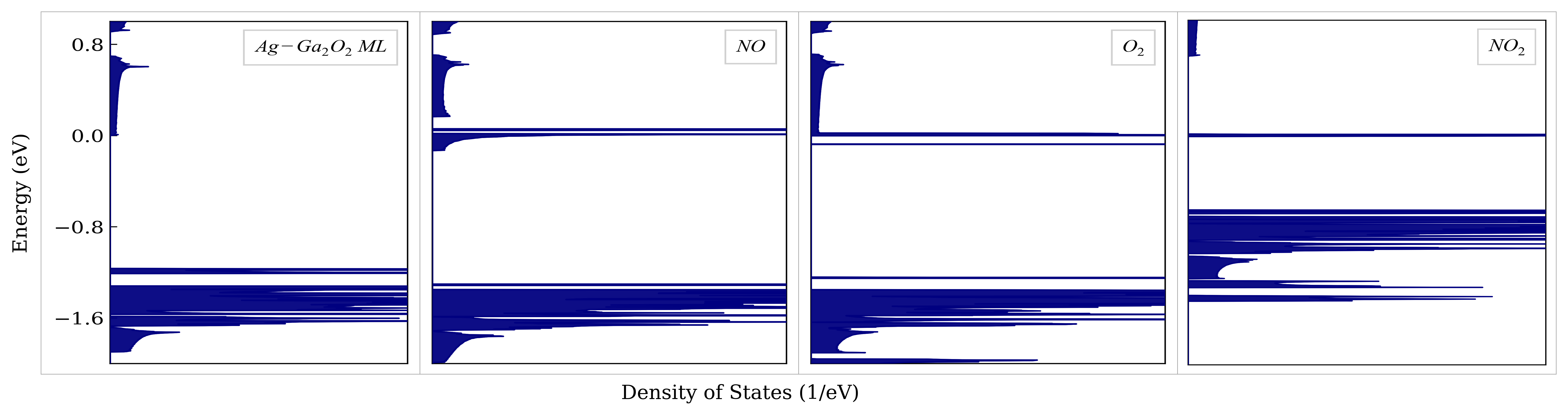}
	\vspace{-0.5cm}
	\caption{Total DOS of the Ag-Ga\textsubscript{2}O\textsubscript{2} ML  and upon adsorption of  target gas molecules. The Fermi level is pinned at $0eV$ in all the panels.}
	\label{fig:DOS_Ag}
\end{figure*}

\noindent The DOS of Ag-Ga\textsubscript{2}O\textsubscript{2} ML exhibits an induced localized filled state located approximately $1.17eV$ below the CBE (see Fig.~\ref{fig:DOS_Ag}). Upon adsorption of NH\textsubscript{3}, CS\textsubscript{2}, H\textsubscript{2}S, and H\textsubscript{2}O, the corresponding electron excitation energies remain largely unchanged, with values of approximately $1.017eV$, $1.17eV$, $1.22eV$ and $1.08eV$, respectively (see Fig.~S17 in SI). Such minor variations in the transition gaps are not expected to induce appreciable changes in the conductivity of the Ag-Ga\textsubscript{2}O\textsubscript{2} ML, thereby yielding no meaningful conductometric response. In the light of this, we calculate the adsorption induced conductivity change factor ($\chi$) to gauge the factor by which the conductivity changes upon adsorption of the molecules~\cite{haque2025predictive, haque2025gas, kalwar2022highly}. The expression for $\chi$ and details of its calculation are provided in Section~7.1 of SI. We note that although there is reasonable modulation in $\chi$ ( Table \ref{table:chi_values}), the large electron excitation energy of $-1.17eV$ keeps the baseline conductivity of Ag-Ga\textsubscript{2}O\textsubscript{2} small. Hence, the change in $\chi$ maybe difficult to detect and prone to modulation due to other issues. Thus, these molecules remain largely invisible and are not expected to produce appreciable and reliable conductometric change in all these four cases. Consequently, these molecules are not considered suitable target analytes for reusable sensing on the Ag-Ga\textsubscript{2}O\textsubscript{2} platform.\\

\noindent CO and SO\textsubscript{2} adsorption modify the electronic structure of Ag-Ga\textsubscript{2}O\textsubscript{2} by increasing the associated excitation energy. The electron excitation energy increases from approximately $1.17eV$ in the Ag-Ga\textsubscript{2}O\textsubscript{2} ML to $1.32eV$ and $1.30eV$ upon CO and SO\textsubscript{2} adsorption, respectively. Although this increase is expected to suppress the conductivity, the relatively large excitation energy already present in the Ag-Ga\textsubscript{2}O\textsubscript{2} limits the magnitude of the adsorption-induced conductivity change. Consequently, the resulting conductometric response is expected to be weak and may be difficult to resolve reliably in a simple resistive sensing setup. Consistent with this interpretation, the calculated $\chi$ values are of the order of $10^{-2}$, indicating conductivity suppression but not necessarily a practically robust sensing signal. Therefore, these molecules cannot be considered as target analytes for Ag-Ga\textsubscript{2}O\textsubscript{2} ML.\\

%Collectively, these results indicate that NH\textsubscript{3}, CO\textsubscript{2}, H\textsubscript{2}O, HF, and CS\textsubscript{2} interact with the  Ag–Ga\textsubscript{2}O\textsubscript{2} surface without introducing gap states or sufficiently perturbing the frontier orbital energies to shift the Fermi level appreciably. Their adsorption is electronically inert from a transport perspective, and they would register negligible signals in a resistive sensing configuration, an important observation for assessing the cross-sensitivity and selectivity of the sensor platform. 
\noindent In sharp contrast, NO ($E_{ads}$=$-0.83eV$) adsorption induces a qualitatively distinct electronic response (see Fig.~\ref{fig:DOS_Ag}). The adsorption of NO induces a partially filled state in close proximity to the CBE. This results in reducing the transition gap to approximately $0.14eV$ ($\approx$6$k_{B}T$ at $T=300K$) drastically. The proximity of this state to the conduction band substantially lowers the carrier excitation threshold, producing an eight-order-of-magnitude enhancement in $\chi$ relative to the Ag-Ga\textsubscript{2}O\textsubscript{2} ML. The resulting conductivity surge upon NO adsorption renders the Ag-Ga\textsubscript{2}O\textsubscript{2} ML a highly sensitive platform for the resistive detection of NO in ambient conditions. Similarly, NO\textsubscript{2} ($E_{ads} = -0.75eV$) adsorption introduces a partially filled state in the vicinity of the Fermi level, located approximately $0.67eV$ ($\approx$26$k_{B}T$ at $T=300K$) below the CBE. Although larger than the corresponding gap in the NO-adsorbed system, this excitation energy gap remains substantially smaller than that of the Ag-Ga\textsubscript{2}O\textsubscript{2} ML and is therefore expected to enhance the conductivity. This behaviour is consistent with the calculated value of $\chi$, which increases by approximately four orders of magnitude upon NO\textsubscript{2} adsorption. Accordingly, both NO and NO\textsubscript{2} emerge as promising target analytes for reusable sensing on the Ag-Ga\textsubscript{2}O\textsubscript{2} ML.\\

\noindent Analogous to NO, adsorption of O\textsubscript{2} ($E_{ads}$=-$0.59eV$) on the Ag–Ga\textsubscript{2}O\textsubscript{2} ML induces a localized filled state at approximately $0.07eV$  ($\approx$3$k_{B}T$ at $T=300K$) below CBE. This near-edge state drives an approximately nine-order-of-magnitude enhancement in $\chi$. Such a substantial enhancement signifies a pronounced increase in electrical conductivity upon O\textsubscript{2} adsorption, comparable to that observed for NO. However, given that O\textsubscript{2} is an ubiquitous constituent of ambient air, such binding and conductometric change with O\textsubscript{2} adsorption, raises a practical concern: under atmospheric conditions, O\textsubscript{2} molecules would competitively occupy the available adsorption sites, effectively passivating the sensor surface and suppressing its response toward other target analytes. This limitation, however, simultaneously defines a compelling niche for the Ag–Ga\textsubscript{2}O\textsubscript{2} ML as a selective O\textsubscript{2} sensor. In oxygen-restricted or controlled-atmosphere environments such as inert-gas gloveboxes, ultra-high vacuum chambers, or hermetically sealed systems used for sensitive physical or chemical experiments, even trace quantities of O\textsubscript{2} represent a critical contamination hazard. The pronounced conductometric response of the Ag–Ga\textsubscript{2}O\textsubscript{2} ML to O\textsubscript{2}, combined with its suitable adsorption energy, positions it as a sensitive and reliable platform for detecting O\textsubscript{2} ingress in such environments, where its cross-sensitivity toward other analytes becomes operationally irrelevant.\\
\vspace{-0.3cm}
\begin{table}[htbp]
	\centering
	\caption{Calculated Adsorption-Induced Conductivity Change Factor ($\chi$) for all favorably adsorbed gas molecules on the Ag-Ga\textsubscript{2}O\textsubscript{2} ML.}
	\begin{tabular}{cc}
			\hline
			Molecules &  Ag Sub \\
			\hline
			NH\textsubscript{3} & 17.84 \\
			NO & 3.62$\times 10^{8}$\\
			SO\textsubscript{2} & 7.48$\times 10^{-2}$ \\
			O\textsubscript{2} & 1.60$\times 10^{9}$ \\
			NO\textsubscript{2} & 1.46$\times 10^{4}$ \\
			H\textsubscript{2}S & 0.35 \\
			H\textsubscript{2}O & 6.40 \\
			CO & 5.08$\times 10^{-2}$ \\
			CS\textsubscript{2} & 0.92 \\
	
			\hline
		\end{tabular}
	\label{table:chi_values}
\end{table}

\noindent \textit{\textbf{Selectivity Analysis for NO and NO\textsubscript{2} under realistic atmospheric conditions}}: Although NO detection may be viable in inert environments, its detection under ambient conditions needs to be analysed due to competition with O\textsubscript{2} for active adsorption sites. For practical detection of NO, the occupancy probability of an active site should be greater or comparable to that of O\textsubscript{2} or other ambient gases. To assess the practical viability of NO detection under ambient conditions, adsorption site occupancy probabilities were computed using a grand canonical adsorption model (Section~3.1, Eq.~S3, SI) under realistic atmospheric conditions i.e. NO at 25 ppm, O\textsubscript{2} at $21\%$, alongside N\textsubscript{2} ($79\%$), H\textsubscript{2}O (5000 ppm), and CO\textsubscript{2} (400 ppm) at 1 atm total pressure, using DFT-calculated adsorption energies~\cite{chen2024transition}. The resulting occupancy probabilities as a function of temperature and NO concentration are shown in Section~7.2.1, Fig.~S21 in SI.\\

\noindent At $300~K$, NO exhibits a site occupancy probability of approximately $55.7\%$, despite its concentration being roughly 840 times lower than that of O\textsubscript{2}. This counterintuitive result is a direct consequence of NO's stronger E\textsubscript{ads} ($-0.83eV$) relative to O\textsubscript{2} ($-0.59eV$). It should be noted that this advantage is sustained around room temperatures but erodes with increasing temperature. At $400~K$, O\textsubscript{2} occupancy probability rises to $~79\%$ while NO falls to $~12\%$, and above $500~K$ empty sites begin to dominate as adsorption weakens exponentially. The concentration variation at $300~K$ identifies a crossover point at approximately 20 ppm below which O\textsubscript{2} dominates site occupancy over NO. Since 25 ppm represents a widely adopted regulatory exposure threshold for NO~\cite{osha_pel, NIOSH_NO}, the sensor operates precisely within the practically relevant detection window under ambient conditions.\\

\noindent The above results indicate that, at or near room temperature, NO (at 25 ppm) and O\textsubscript{2} (at 21\%) would occupy the active sites with equal probabilities. Consequently, in the absence of NO, the adsorption sites are expected to be predominantly occupied by O\textsubscript{2}. However, upon exposure to NO at concentrations of 20–25 ppm, which is the permissible exposure limit (PEL) for NO~\cite{osha_pel, NIOSH_NO}, about half of these sites becomes occupied by NO at $300K$. Therefore, practical NO detection under ambient conditions relies on the existence of a discernible conductometric difference between the O\textsubscript{2}-adsorbed ML and the co-adsorbed O\textsubscript{2}/NO ML. To assess this possibility, it is necessary to examine the electronic structure of the monolayer under simultaneous adsorption of O\textsubscript{2} and NO and evaluate whether the resulting conductance modulation remains sufficiently distinct for reliable NO detection in ambient environments. Hence, to further investigate the practical implications of O\textsubscript{2}'s strong conductometric response under ambient conditions, a DFT simulation was performed by placing both O\textsubscript{2} and NO simultaneously on an 8×4×1 Ga\textsubscript{2}O\textsubscript{2} supercell incorporating two Ag substitution sites, one for each molecule with sufficient intermolecular separation to preclude direct molecule-molecule interaction (see Section~7.2, Fig.~S19 in SI). The resulting DOS reveals a qualitatively significant departure from that of the isolated O\textsubscript{2} case: the localized state previously observed at the CBE is absent, and instead a partially filled localised state emerges with the excitation energy of $\approx$ $0.22eV$, closely mirroring the electronic signature of isolated NO adsorption (see Section~7.2, Fig.~S20 in SI). Crucially, the suppression of the O\textsubscript{2} induced state at the CBE in the presence of NO is a favorable outcome from a sensing perspective. Hence, it is expected that in the ambient environment around the room temperature; Ag-Ga\textsubscript{2}O\textsubscript{2} ML will show large conductivity due to adsorption of sufficient O\textsubscript{2} molecule. However, once exposed to NO, the conductivity would drop appreciably due to the increase in the electron excitation energy to $0.22eV$. It should be noted that although NO adsorption increases the electron excitation energy to $0.22eV$, this value corresponds to only $\approx$9$k_{B}T$ at room temperature. Therefore, while a measurable reduction in conductivity is expected, the conductivity is unlikely to be suppressed by several orders of magnitude. \\

\noindent Collectively, both the site occupancy analysis and co-adsorption DOS consistently establish that the Ag–Ga\textsubscript{2}O\textsubscript{2} ML retains selective sensitivity toward NO under ambient conditions, with NO's stronger adsorption energy serving as the decisive factor in both surface site competition and electronic structure determination.\\

\noindent \noindent \textit{\textbf{Assessment of NO\textsubscript{2} sensing under ambient conditions}}: Similar to NO, the practical viability of NO\textsubscript{2} sensing under ambient conditions was evaluated by computing occupancy probabilities for active sites. The concentration of NO\textsubscript{2} was taken as 25 ppm and 40 ppm. Despite the favorable adsorption energy of NO\textsubscript{2} ($-0.75eV$), the substantially higher concentration of O\textsubscript{2} results in preferential occupation of the active sites by O\textsubscript{2}. At 25 ppm of NO\textsubscript{2}, its calculated site occupancy probability is $\sim19\%$ while for O\textsubscript{2}, it is $\sim81\%$. Increasing the NO\textsubscript{2} concentration to 40 ppm raises the NO\textsubscript{2} occupancy to only $\sim27\%$. In fact, the calculated crossover concentration, at which NO\textsubscript{2} and O\textsubscript{2} exhibit comparable site occupancies, is approximately 9400 ppm, far exceeding environmentally relevant NO\textsubscript{2} levels. This further confirms that competitive adsorption by O\textsubscript{2} is expected to suppress practical NO\textsubscript{2} sensing under ambient conditions. %Consequently, O\textsubscript{2} remains the dominant surface species under realistic atmospheric conditions, limiting the practical accessibility of the active sites for NO\textsubscript{2} detection.
Nevertheless, the favorable adsorption energy and conductometric response of NO\textsubscript{2} suggest that its detection may still be feasible under O\textsubscript{2} controlled or O\textsubscript{2}-deficient environments, where competitive occupation by atmospheric O\textsubscript{2} is substantially reduced. Similar to the NO co-adsorption case, a simulation was performed by simultaneously placing O\textsubscript{2} and NO\textsubscript{2} on an 8×4×1 Ga\textsubscript{2}O\textsubscript{2} supercell containing two Ag substitution sites, one for each molecule, while maintaining sufficient intermolecular separation to eliminate direct molecule–molecule interactions (see Section~7.3, Fig.~S22 in SI). The resulting DOS exhibits a qualitatively distinct electronic structure compared to the isolated O\textsubscript{2}-adsorbed system. In particular, the localized state previously located at the CBE is no longer present. Instead, a partially filled state emerges, giving rise to an excitation energy of approximately $0.84eV$ to CB. Notably, this electronic signature closely resembles that of the isolated NO\textsubscript{2}-adsorbed system (see Section~7.3, Fig.~S23 in SI), suggesting that despite the dominant occupancy of O\textsubscript{2}, the electronic response of the co-adsorbed configuration is governed primarily by the presence of NO\textsubscript{2}. The detailed occupancy probability calculations and co-adsorption analysis are provided in Section~7.3 of SI, respectively. Nevertheless, despite its intrinsic selectivity toward NO\textsubscript{2}, competitive occupation by atmospheric O\textsubscript{2} limits the practical viability of the Ag-Ga\textsubscript{2}O\textsubscript{2} ML for ambient NO\textsubscript{2} sensing.

\subsection{Ti and Pt Substitution}
\label{subsec:Ti_sub}
\begin{figure*}[htbp]
	\centering
	
	\begin{minipage}[c]{0.50\textwidth}
		\centering
		\scriptsize
		\setlength{\tabcolsep}{6pt}
		\renewcommand{\arraystretch}{1.3}
		
		\begin{tabular}{lcccc}
			\hline
			Molecule & $E_{ads}$ (eV) & H (\AA) & q (e) & $\tau$ (s) \\
			\hline
			NH\textsubscript{3}  & -6.57 & 2.28 & 0.11 & $2.35 \times 10^{98}$ \\
			NO                  & -6.63 & 1.92 & 0.30 & $2.39 \times 10^{99}$ \\
			SO\textsubscript{2} & -5.57 & 2.97 & 0.17 & $3.73 \times 10^{81}$ \\
			O\textsubscript{2}  & -8.15 & 1.74 & 0.66 & $8.20 \times 10^{124}$ \\
			NO\textsubscript{2} & -7.57 & 1.89 & 0.52 & $1.48 \times 10^{115}$ \\
			H\textsubscript{2}S & -5.99 & 2.80 & 0.08 & $4.24 \times 10^{88}$ \\
			H\textsubscript{2}O & -5.84 & 2.69 & 0.007 & $1.28 \times 10^{86}$ \\
			CO                  & -5.74 & 2.52 & 0.008 & $2.68 \times 10^{84}$ \\
			CO\textsubscript{2} & -5.45 & 3.34 & 0.02 & $3.60 \times 10^{79}$ \\
			CS\textsubscript{2} & -5.80 & 3.12 & 0.01 & $2.73 \times 10^{85}$ \\
			HF                  & -5.83 & 2.36 & 0.008 & $8.70 \times 10^{85}$ \\
			\hline
		\end{tabular}
		
		\vspace{2.0em}
		{\normalsize\textbf{(a)}}
	\end{minipage}
	\hfill
	\begin{minipage}[c]{0.48\textwidth}
		\centering
		\includegraphics[width=0.95\linewidth]{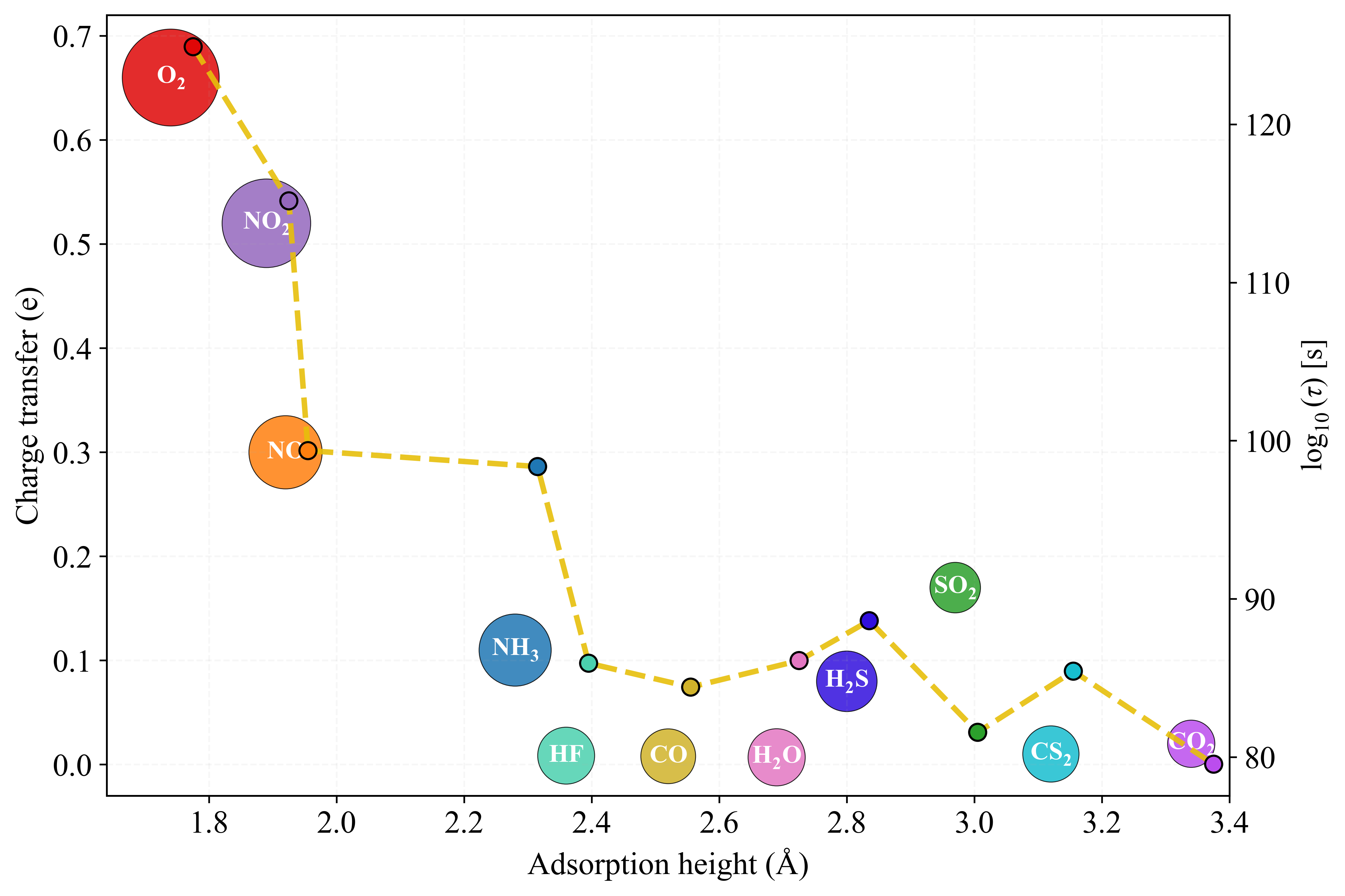}
		
		\vspace{0.3em}
		\textbf{(b)}
	\end{minipage}
	
	\caption{(a) Table VI: Adsorption parameters of the investigated gas molecules on the Ti-Ga\textsubscript{2}O\textsubscript{2} ML, including adsorption energy ($E_{ads}$), adsorption height (H), charge transfer (q), and recovery time ($\tau$). (b) Graphical summary of the adsorption characteristics of gas molecules on the Ti-Ga\textsubscript{2}O\textsubscript{2} monolayer. The centers of the larger bubbles are referenced to the X and primary Y-axes, representing the adsorption height and charge transfer, respectively. The area of each bubble is proportional to the magnitude of the adsorption energy. The smaller markers linked by the dashed line indicate the corresponding recovery times, referenced to the secondary Y-axis. Recovery times are reported on a logarithmic scale.}
	\label{fig:Ti_summary}
\end{figure*}

\textit{\textbf{Ti-substituted System}}: The Ti-substituted ML (Ti-Ga\textsubscript{2}O\textsubscript{2}) exhibits the most aggressive adsorption characteristics among all substitutional configurations examined in this study. All eleven analytes display adsorption energies far exceeding the $-1.5eV$ irreversibility threshold, ranging from a maximum of $-5.45eV$ for CO\textsubscript{2} to a minimum of $-8.15eV$ for O\textsubscript{2}, with very prolonged recovery times, (see Fig.~\ref{fig:Ti_summary}). This universally extreme binding behavior unambiguously classifies the Ti-Ga\textsubscript{2}O\textsubscript{2} ML as a broad-spectrum chemical scavenger, capable of permanently immobilizing all target analytes regardless of their chemical nature or molecular polarity.\\

\noindent Examination of adsorption heights and charge transfer reveals two dominant interaction characteristics at the Ti site. The first is observed for O\textsubscript{2} (H=1.74Å, Q=0.66e), NO\textsubscript{2} (1.89Å, 0.52e), and NO (1.92Å, 0.30e), where low adsorption heights and substantial charge transfer indicate strong adsorption accompanied by significant electron-density transfer from the Ti-substituted surface to the adsorbed molecules. The significant charge transfer upon adsorption is expected to strengthen the electrostatic attraction between the adsorbate and the Ti-substituted surface, which is reflected in the relatively short adsorption distances. The relaxed geometries further exhibit noticeable molecular reorientation and, in selected cases, mild structural distortion consistent with chemisorptive behaviour (see Fig.~S19 in SI).\\

\noindent The second interaction characteristic is observed for H\textsubscript{2}O, CO, CO\textsubscript{2}, CS\textsubscript{2}, and HF, which exhibit negligible net charge transfer ($\leq 0.02e$) despite very strong adsorption energies ($-5.45$ to $-5.84eV$). Together with the relaxed geometries, this indicates that their binding is not governed by net ionic charge transfer, but rather by strong covalent-type interaction between molecular orbitals and Ti $d$-states. The low net charge transfer may arise from compensated donation and back-donation processes, leading to substantial charge redistribution without a large net electron transfer~\cite{kalwar2022highly}. NH\textsubscript{3}, SO\textsubscript{2}, and H\textsubscript{2}S exhibit intermediate behavior, with moderate charge transfer (0.11e, 0.17e, and 0.08e, respectively), suggesting a mixed nature of interaction.\\

\noindent Given the universally irreversible adsorption nature, DOS analysis is not pursued for the Ti–Ga\textsubscript{2}O\textsubscript{2} ML. The extreme adsorption energies across all eleven gas molecules spanning ambient, toxic, and corrosive species alike unambiguously establish its role as a broad-spectrum chemical scavenger. Hence, the potential utility of Ti–Ga\textsubscript{2}O\textsubscript{2} ML may lie in creating and maintaining chemically inert or ultra-clean environments where complete removal of gas-phase species is desired.\\

\begin{figure*}[htbp]
	\centering
	
	\begin{minipage}[c]{0.50\textwidth}
		\centering
		\scriptsize
		\setlength{\tabcolsep}{6pt}
		\renewcommand{\arraystretch}{1.3}
		
		\begin{tabular}{lcccc}
			\hline
			Molecule & $E_{ads}$ (eV) & H (\AA) & q (e) & $\tau$ (s) \\
			\hline
		NH\textsubscript{3}  & -6.67 & 1.63 & -0.23 & $1.12 \times 10^{100}$ \\
		NO                  & -6.48 & 1.30 & -0.24 & $7.23 \times 10^{96}$ \\
		SO\textsubscript{2} & -5.88 & 2.56 & 0.005 & $6.02 \times 10^{86}$ \\
		O\textsubscript{2}  & -5.78 & 2.58 & -0.005 & $1.26 \times 10^{85}$ \\
		NO\textsubscript{2} & -6.16 & 1.56 & 0.23 & $3.04 \times 10^{91}$ \\
		H\textsubscript{2}S & -6.24 & 1.99 & -0.23 & $6.72 \times 10^{92}$ \\
		H\textsubscript{2}O & -6.15 & 1.66 & -0.08 & $2.07 \times 10^{91}$ \\
		CO                  & -6.10 & 1.50 & -0.06 & $2.99 \times 10^{90}$ \\
		CO\textsubscript{2} & -5.82 & 2.51 & 0.01 & $5.91 \times 10^{85}$ \\
		CS\textsubscript{2} & -5.94 & 2.72 & -0.009 & $6.13 \times 10^{87}$ \\
		HF                  & -5.77 & 2.26 & 0.004 & $8.55 \times 10^{84}$ \\
			\hline
		\end{tabular}
		
		\vspace{2.0em}
		{\normalsize\textbf{(a)}}
	\end{minipage}
	\hfill
	\begin{minipage}[c]{0.48\textwidth}
		\centering
		\includegraphics[width=0.95\linewidth]{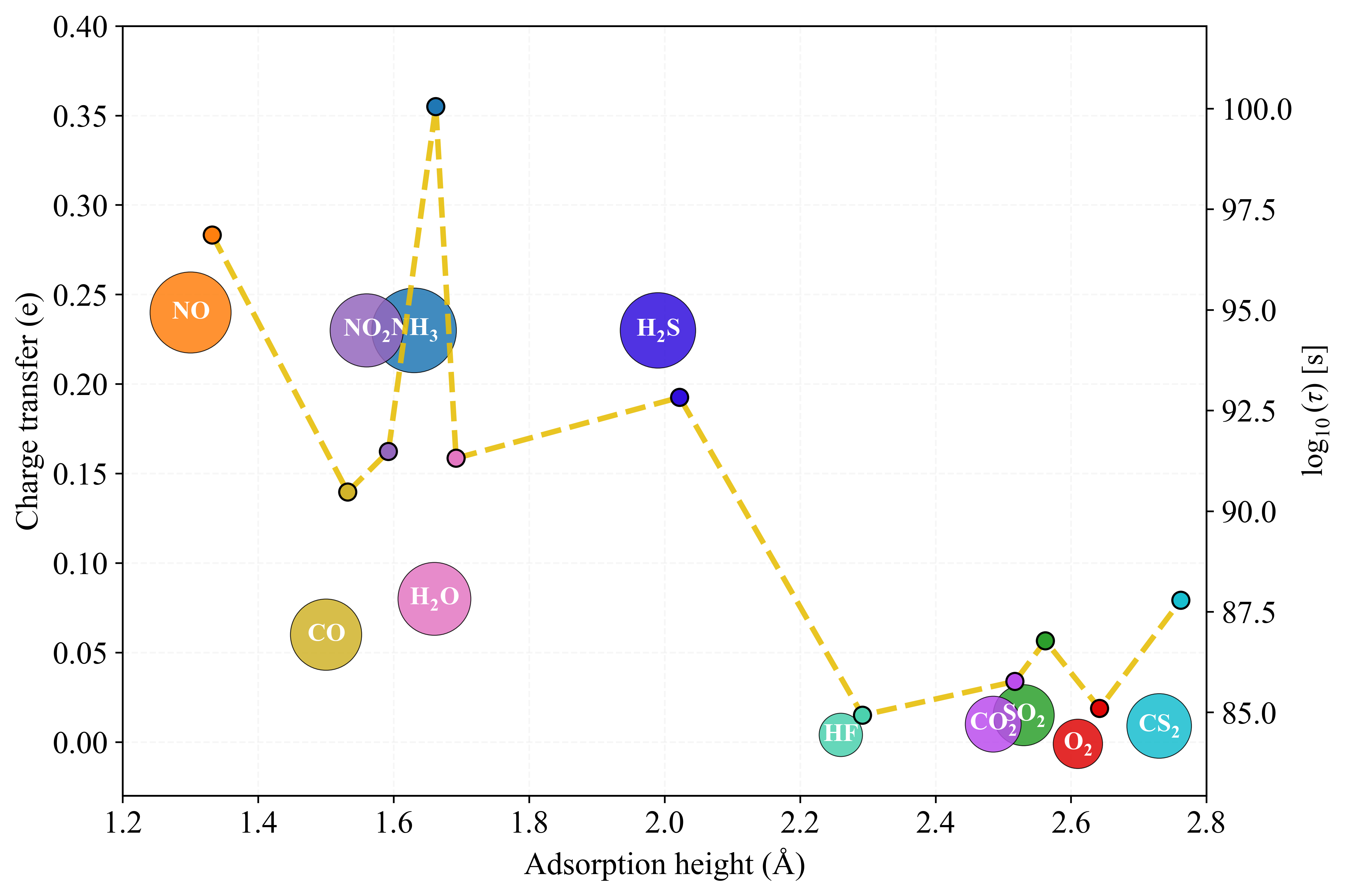}
		
		\vspace{0.3em}
		\textbf{(b)}
	\end{minipage}
	
	\caption{(a) Table VII: Adsorption parameters of the investigated gas molecules on the Pt-Ga\textsubscript{2}O\textsubscript{2} ML, including adsorption energy ($E_{ads}$), adsorption height (H), charge transfer (q), and recovery time ($\tau$). (b) Graphical summary of the adsorption characteristics of gas molecules on the Pt-Ga\textsubscript{2}O\textsubscript{2} monolayer. The centers of the larger bubbles are referenced to the X and primary Y-axes, representing the adsorption height and charge transfer, respectively. The area of each bubble is proportional to the magnitude of the adsorption energy. The smaller markers linked by the dashed line indicate the corresponding recovery times, referenced to the secondary Y-axis. Recovery times are reported on a logarithmic scale.}
	\label{fig:Pt_summary}
\end{figure*}

\noindent \textit{\textbf{Pt-substituted System}}: The Pt-substituted ML (Pt-Ga\textsubscript{2}O\textsubscript{2}) mirrors same behaviour as the Ti–Ga\textsubscript{2}O\textsubscript{2} ML in its broad-spectrum scavenging behavior, with all analytes exhibiting adsorption energies ranging from a maximum of $-5.77{eV}$ to a minimum of $-6.67{eV}$ for HF and NH\textsubscript{3} respectively, and hence, extremely long recovery times, (see Fig.~\ref{fig:Pt_summary}). This firmly establishes irreversible molecular capture across the entire panel of target gases. However, the underlying adsorption mechanism differs fundamentally from that of the Ti site.\\

\noindent Unlike Ti, where two distinct interaction characteristics were identified through substantial charge transfer values, the Pt site is characterized by near-universally low charge transfer magnitudes across all molecules, with values rarely exceeding $0.1e$. This suggests that the binding on the Pt site is predominantly covalent in character, driven by orbital hybridization between the adsorbate and the Pt d-states rather than ionic charge redistribution. The adsorption heights, ranging from 1.30\AA~for NO to 2.72\AA~for CS\textsubscript{2}, are consistent with strong orbital overlap supporting this hybridization-dominated picture. The Pt-Ga\textsubscript{2}O\textsubscript{2} ML is therefore also conclusively identified as a broad-spectrum chemical scavenger, with the distinction that its capture mechanism is predominantly covalent and hybridization-driven.\\
\begin{figure}[!ht]
	\centering
	\includegraphics[width=0.84\textwidth]{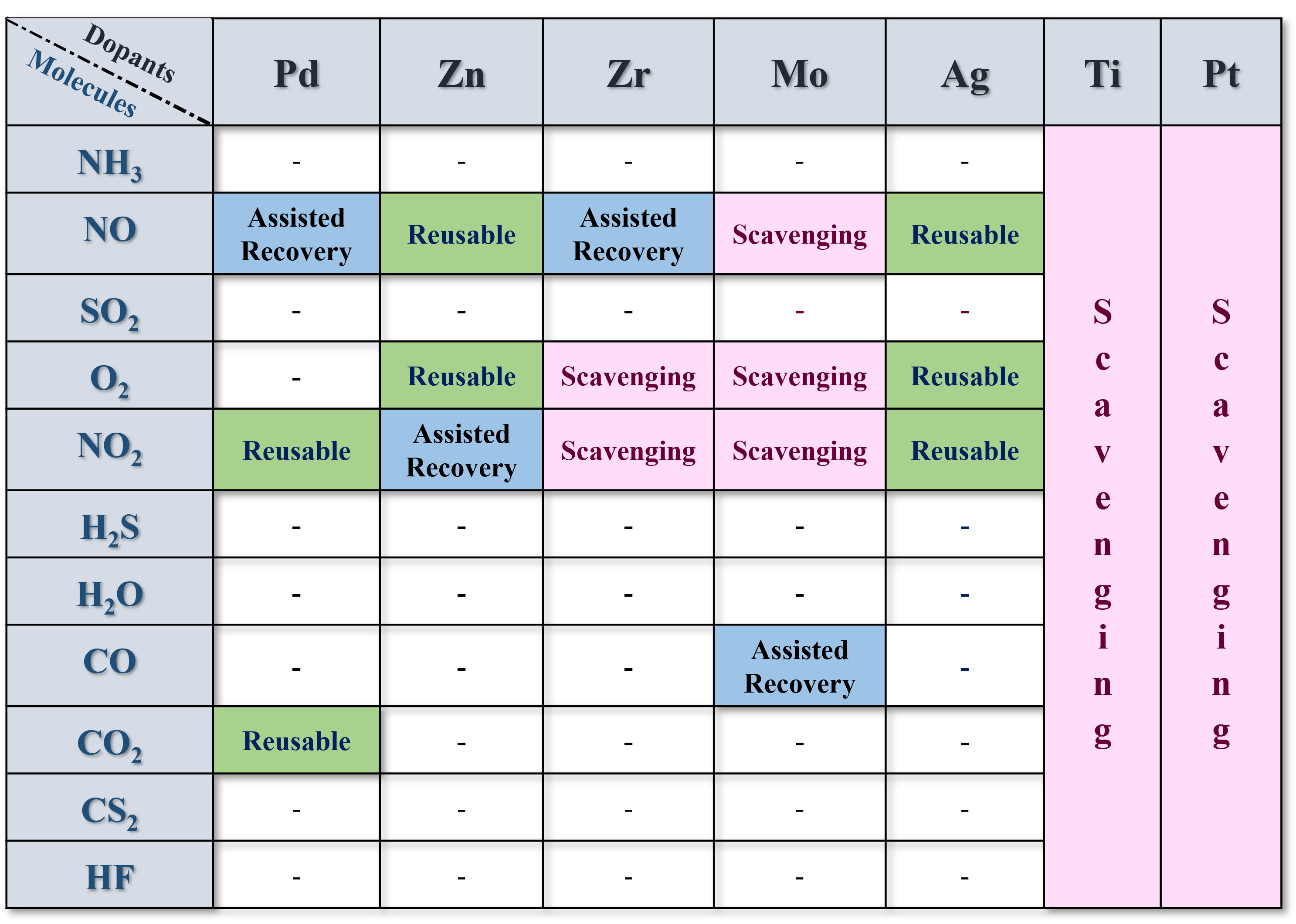}
	\caption{Summary of the sensing and scavenging characteristics of the investigated transition metal-substituted Ga\textsubscript{2}O\textsubscript{2} monolayer toward the studied gas molecules. The color coding identifies reusable sensing applications (green), sensing applications requiring external assistance for recovery (blue), and scavengers exhibiting effectively irreversible adsorption (pink). Blank cells correspond to systems that do not exhibit practically relevant sensing or scavenging performance under the adopted screening criteria.}
	\label{fig:summary}
\end{figure}

\noindent  Section~11 of SI provides a comparative summary of literature-reported adsorption energies and recovery times for NO, NO\textsubscript{2}, O\textsubscript{2}, CO, and CO\textsubscript{2} on various 2D materials. The complete adsorption-energy landscape of the investigated TM-substituted Ga\textsubscript{2}O\textsubscript{2} monolayers is presented as a heat map in Fig.~S26 in SI. The resulting sensing and scavenging functionalities, classified according to the adopted screening criterion, are summarized in Fig.~\ref{fig:summary}.

\section{Conclusion} 
\label{conclusion}

In summary, this work presents a systematic first-principles investigation of transition metal substituted Ga\textsubscript{2}O\textsubscript{2} monolayer as platforms for gas sensing and molecular scavenging, motivated by the limited analyte range and absence of irreversible capture capability in the pristine Ga\textsubscript{2}O\textsubscript{2} surface. Across seven dopant species and eleven target gas molecules, substitutional doping is shown to dramatically expand the functional landscape of the Ga\textsubscript{2}O\textsubscript{2} ML, from selective room-temperature resistive sensing to broad-spectrum irreversible molecular capture. Pd-Ga\textsubscript{2}O\textsubscript{2} supports reusable sensing of NO\textsubscript{2} and CO\textsubscript{2}, with NO sensing remaining accessible through assisted-recovery operation. Zn-Ga\textsubscript{2}O\textsubscript{2} further expands the sensing functionality of the substituted ML, enabling reusable detection of NO and O\textsubscript{2}, while NO\textsubscript{2} lies within the assisted-recovery regime. Beyond these reusable sensing platforms, Zr-Ga\textsubscript{2}O\textsubscript{2} exhibits sensing potential toward NO within the assisted-recovery regime, while selectively scavenging O\textsubscript{2} and NO\textsubscript{2}. In addition, Mo-Ga\textsubscript{2}O\textsubscript{2} displays analogous sensing behavior toward CO and additionally functions as a selective scavenger for NO, NO\textsubscript{2}, and O\textsubscript{2}. Among all investigated systems, Ag-Ga\textsubscript{2}O\textsubscript{2} produces the most pronounced conductometric response, yielding an approximately eight and nine -order-of-magnitude enhancement in electrical conductivity upon NO and O\textsubscript{2} adsorption at room temperature. In addition, Ag-Ga\textsubscript{2}O\textsubscript{2} exhibits reusable sensing potential toward O\textsubscript{2} and NO\textsubscript{2}. Additional analysis of selectivity in the ambient environment demonstrate that Pd-Ga\textsubscript{2}O\textsubscript{2} and Ag-Ga\textsubscript{2}O\textsubscript{2} remain selective towards NO while Zn-Ga\textsubscript{2}O\textsubscript{2} remains selective towards NO\textsubscript{2} under normal atmospheric conditions. At the extreme end, Ti and Pt drive universally irreversible molecular capture, a capability that eluded the pristine Ga\textsubscript{2}O\textsubscript{2} surface entirely. Beyond the scope of the present work, the exploration of additional dopant species and the effects of adatom functionalization remain promising avenues for further expanding the functional versatility of the Ga\textsubscript{2}O\textsubscript{2} monolayer. These directions remain the subject of future investigation.

\section*{Acknowledgements}

A.A.H. acknowledges the Ministry of Education, Govt. of India, for the Prime Minister's Research Fellowship (PMRF). The authors acknowledge National Supercomputing Mission (NSM) for providing computing resources of ‘PARAM Shakti’ at IIT Kharagpur, implemented by C-DAC and supported by the Ministry of Electronics and Information Technology (MeitY) and Department of Science and Technology (DST), Government of India.

\section*{Supporting information}

Supplementary Information (SI) available: AIMD plots, Optimized adsorption configurations, DOS plots, charge density difference isosurfaces, conductivity change factor ($\chi$) calculations, competitive adsorption analysis, and additional computational details.

\bibliographystyle{ieeetran}
\bibliography{myref}

\end{document}